\documentclass[]{jfm}

\usepackage{graphicx}
\usepackage{newtxtext}
\usepackage{ulem}
\usepackage{newtxmath}
\usepackage{natbib}
\usepackage{hyperref}
\usepackage{color}
\hypersetup{
    colorlinks = true,
    urlcolor   = blue,
    citecolor  = blue,
}

\newcommand{\RomanNumeralCaps}[1]
\linenumbers

\title{Linear instability in thermally-stratified quasi-Keplerian flows}

\author{Dongdong Wan\aff{1}, 
Rikhi Bose\aff{2},
  Mengqi Zhang\aff{1,3} \corresp{\email{mpezmq@nus.edu.sg}}
 \and Xiaojue Zhu\aff{2}  \corresp{\email{zhux@mps.mpg.de}}  } 

\affiliation{\aff{1}Department of Mechanical Engineering, National University of Singapore, 9 Engineering Drive 1, 117575, Singapore
\aff{2} Max Planck Institute for Solar System Research, 37077 G\"ottingen, Germany
\aff{3} NUS Research Institute (NUSRI) in Suzhou, No. 377 Linquan Street, Suzhou, Jiangsu, 215123, China
}

\begin{document}
\maketitle

\begin{abstract}
Quasi-Keplerian flow, a special regime of Taylor-Couette co-rotating flow, is of great astrophysical interest for studying angular momentum transport in accretion disks. The well-known magnetorotational instability (MRI) successfully explains the flow instability and generation of turbulence in certain accretion disks, but fails to account for these phenomena in protoplanetary disks where magnetic effects are negligible. Given the intrinsic decrease of the temperature in these disks, we examine the effect of radial thermal stratification on 3-D global disturbances in linearised quasi-Keplerian flows under radial gravitational acceleration mimicking stellar gravity.
Our results show a thermohydrodynamic linear instability for both axisymmetric and non-axisymmetric modes across a broad parameter space of the thermally-stratified quasi-Keplerian flow.
Generally, decreasing Richardson or Prandtl numbers stabilises the flow, while a reduced radius ratio destabilises it. This work also provides a quantitative characterisation of the instability.
At low Prandtl numbers $Pr$, we observe a scaling relation of the linear critical Taylor number $Ta_c\propto$$Pr^{-6/5}$. Extrapolating the observed scaling to high $Ta$ and low $Pr$ may suggest the relevance of the instability to accretion disks. Moreover, even slight thermal stratification, characterised by a low Richardson number, can trigger the flow instability with a small axial wavelength.
These findings are qualitatively consistent with the results from a traditional local stability analysis based on short wave approximations. Our study refines the thermally-induced linearly-unstable transition route in protoplanetary disks to explain angular momentum transport in dead zones where MRI is ineffective.
\end{abstract}


\section{Introduction}
\label{sec:introduction}

Accretion disks are thin reservoirs of dispersed material (e.g., gas and dust) orbiting massive central objects, such as growing galaxies, planets, stars, and black holes \citep{Ji2013Angular}. Their dynamics has been traditionally modelled by a type of Taylor-Couette (TC) flow, where the inner cylinder rotates faster than the outer one, but the angular momentum increases radially outward, i.e., $\omega_i^*>\omega_o^*$ and $\omega_i^* r_i^{*2}<\omega_o^* r_o^{*2}$ \citep{Dubrulle2005Stability,Avila2012Stability,Ji2013Angular}. This type of TC flow is known as quasi-Keplerian flow, as depicted in figure \ref{Fig:schematic_flowregime}. Based on the TC flow model, the current study aims to better elucidate the angular momentum transport mechanism in accretion disks.

For the central object within a disk to continuously gain mass from the surrounding material, the ``fluid particles'' must lose angular momentum to fall inward under gravity. Observed accretion rates cannot be fully explained by angular momentum transport due to the molecular viscosity alone \citep{Pringle1981Accretion,Grossmann2016High}, prompting researchers to explore alternative mechanisms. \cite{Shakura1973Black} first proposed that turbulence could play a crucial role in this process. Since then, considerable effort has been devoted to searching for flow instability mechanisms that could lead to turbulence in quasi-Keplerian flows. Nevertheless, hydrodynamic considerations, excluding other magnetic or thermal effects, indicate that  quasi-Keplerian flows are linearly stable to axisymmetric disturbances, according to Rayleigh's criterion \citep{Rayleigh1917Dynamics}. Recent calculations for non-axisymmetric disturbances also suggest linear stability in astrophysically interesting regimes \citep{Deguchi2017Linear}.
This hydrodynamic linear stability raises the question of how turbulence arises to facilitate the radial transport of angular momentum in accretion disks.

\subsection{Magnetorotational instability and hydrodynamic subcritical nonlinear instability}\label{sec:MRI_subcritical}

A significant milestone in addressing the angular momentum transport problem was reached by \cite{Balbus1991Powerful}, who introduced the magnetorotational instability (MRI) concept, originally proposed by \cite{Velikhov1959Stability}, into the astrophysics community. They discovered that a quasi-Keplerian flow could be linearly unstable even in the presence of very weak magnetic fields. Over 30 years later, MRI was successfully detected in a laboratory Taylor-Couette flow experiment \citep{Wang2022Identification,Wang2022Observation}. This success supports using quasi-Keplerian flow at laboratory scales to study the physical processes in accretion disks on astronomical scales. However, the MRI diagram struggles to explain the angular momentum transport in protoplanetary or protostellar disks. This is because in these large, cool and dusty disks, the ionisation level in certain zones (dubbed ``dead zones'') is so low that magnetising the fluid particles is difficult or impossible, rendering this diagram inapplicable \citep{Gammie1996Layered,Fleming2003Local,Armitage2011Dynamics,Held2018Hydrodynamic}.

With the magnetic field being inoperative, researchers have explored hydrodynamic instabilities with other effects to account for radial angular momentum transport in dead zones. One line of research has focused on subcritical nonlinear instability, as linear stability does not necessarily ensure nonlinear stability \citep{Grossmann2000Onset,Eckhardt2007Turbulence}. Unfortunately, numerical simulations and laboratory experiments have shown no signs of subcritical transition in quasi-Keplerian flows, even at Reynolds numbers as high as millions \citep{Ji2006Hydrodynamic,Ji2013Angular,Ostilla2014Turbulence,Edlund2014Nonlinear,Shi2017Hydrodynamic}. Given that typical Reynolds numbers in accretion disks can reach $10^{12}\sim10^{14}$ \citep{Ji2006Hydrodynamic,Fromang2019Angular}, much higher than currently explored values, we cannot entirely dismiss the possibility of subcritical transition therein. Another line of research has examined the thermal effects on flow stability/instability in accretion disks, which is the direction the present work follows, to be reviewed below.

\subsection{Hydrodynamic instability driven by thermal effects}\label{sec:research_efforts}

The thermal structure of accretion disks is complex and non-uniform, leading to intricate thermodynamics that shapes the disks' structure \citep{lesur2022}. For instance, in protoplanetary disks, numerical modelling and calculations based on astrophysical observations \citep{Chiang1997Spectral,Fromang2019Angular} showed that the mid-plane is relatively cool (due to radiation to space) and sandwiched between two hot surface layers (due to irradiation by the central star). These studies also demonstrated that the mid-plane temperature decreases with increasing distance from the central star, indicating radial thermal stratification.

Considering the non-uniform thermal structure, various instabilities due to the thermal effect have been identified in the modelling and analyses of accretion disks. These include the two-dimensional (2-D) subcritical baroclinic instability in the $r$-$\theta$ plane due to non-axisymmetric disturbances \citep{Lesur2010Subcritical}, the three-dimensional (3-D) linear convective overstability due to axisymmetric disturbances \citep{Klahr2014Convective}, and the axisymmetric 2-D linear vertical shear instability in the $r$-$z$ plane \citep{Nelson2013Linear}. For a comprehensive review of thermally-related instability mechanisms and others, including gravitational instability, zombie vortex instability, and stratorotational instability, the reader is referred to \cite{Fromang2019Angular}.
The richness of these instabilities demonstrates that many physical effects, additional to the magnetic effects, can destabilise the canonical Keplerian flow. Nevertheless, their effectiveness depends on the disk structures and the thermal timescales of the gases \citep{lesur2022}. 
Among these instabilities, the baroclinic instability, caused by radial buoyancy forces, is a nonlinear mechanism. The linear convective overstability has been analysed for inviscid flows within a local planar shearing box. Following the shearing box approximation adopted by \cite{Klahr2014Convective}, \cite{Lyra2014Convective} confirmed the linear convective overstability and highlighted the need to investigate its operation in ``global models''.
To our knowledge, 3-D ``global'' linear instability concerning the thermodynamics of accretion disks, especially in the radial direction, has rarely been reported, which is the focus of the present work. By ``global'', we specifically mean that the boundary conditions at the two rotating cylinders are directly incorporated in the problem formulation based on the quasi-Keplerian flow model in a Taylor-Couette geometry. It should be mentioned that the above-mentioned stratorotational instability \citep{Yavneh2001Non} is a linear instability identified in TC flow, but it requires stable vertical stratification and the separation between cylinder walls must be sufficiently close for the instability to be effective. 

\subsection{The position and structure of the current work}

In light of the efforts to uncover possible hydrodynamic instabilities without magnetic fields, this study incorporates radial thermal stratification along with stellar gravity in the modelling of accretion disks using Taylor-Couette flow with hot inner and cold outer walls in the quasi-Keplerian regime. The objective is to identify possible 3-D thermo-hydrodynamic linear instability mechanisms that may operate in the thermal TC flow. In our flow system, the thermal gradient aligns with the direction of stellar gravity.

While it is well-known that thermal buoyancy is destabilising and quasi-Keplerian shear is strongly stabilising \citep{Ji2006Hydrodynamic}, to the best of our knowledge, the quantitative aspects of their interplay in the considered flow have not been thoroughly detailed in the literature.
Similar flow configurations exist but differ from our system, such as rotating Rayleigh-B{\'e}nard convection in a closed cylindrical container \citep{Ecke2023Turbulent} and centrifugal TC flows with hot outer and cold inner cylinders \citep{Becker1962Influence,Ali1990Stability,Meyer2015Effect,Kang2017Radial,Jiang2020Supergravitational,Meyer2021Stability,Wang2022Effects,Jiang2022Experimental}, where gravitational effects are ignored. To illustrate these differences, we compare our flow configuration with a representative work \cite{Meyer2021Stability} in table \ref{Tab:compare_Meyer2021PRF} in Appendix \ref{app:two_studies}. Additionally, our 3-D ``global'' thermally-driven linear instability also contrasts with the local shearing-box approximations and short wave approximations used in prior works such as \cite{Klahr2014Convective}, \cite{Lyra2014Convective}, \cite{Latter2016Convective}, and \cite{Held2018Hydrodynamic}. A comparison of our problem setting with \cite{Klahr2014Convective} is provided in table \ref{Tab:compare_Klahr2014} in Appendix \ref{app:two_studies}.

The rest of the paper is organised as follows. Section \ref{sec:problem_formulation} describes the flow configuration, governing equations, and linear stability analysis as an eigenvalue problem. The numerical method is introduced in section \ref{sec:methods_validations}. Section \ref{sec:results} presents and discusses the results, characterising the linear instability at moderate values of the parameters (including Taylor number, Richardson number, Prandtl number, and radius ratio) to examine their effects on the linear instability and the scaling relation. We also demonstrate that the linear instability persists within extreme parameter ranges relevant to accretion disks, even with weak thermal buoyancy effects, highlighting the effectiveness of this thermal instability. Comparisons to previous works are made where possible. Section \ref{sec:conclusion} concludes the paper. Appendices summarise the comparison to previous works and the influence of the gravitational acceleration profile on the linear instability.
 
\section{Problem formulation}\label{sec:problem_formulation}

\subsection{Flow configuration and governing equations}\label{sec:configuration_equations}

\begin{figure}
	\centering{
	\includegraphics[width=0.45\textwidth,trim= 100 0 100 30,clip]{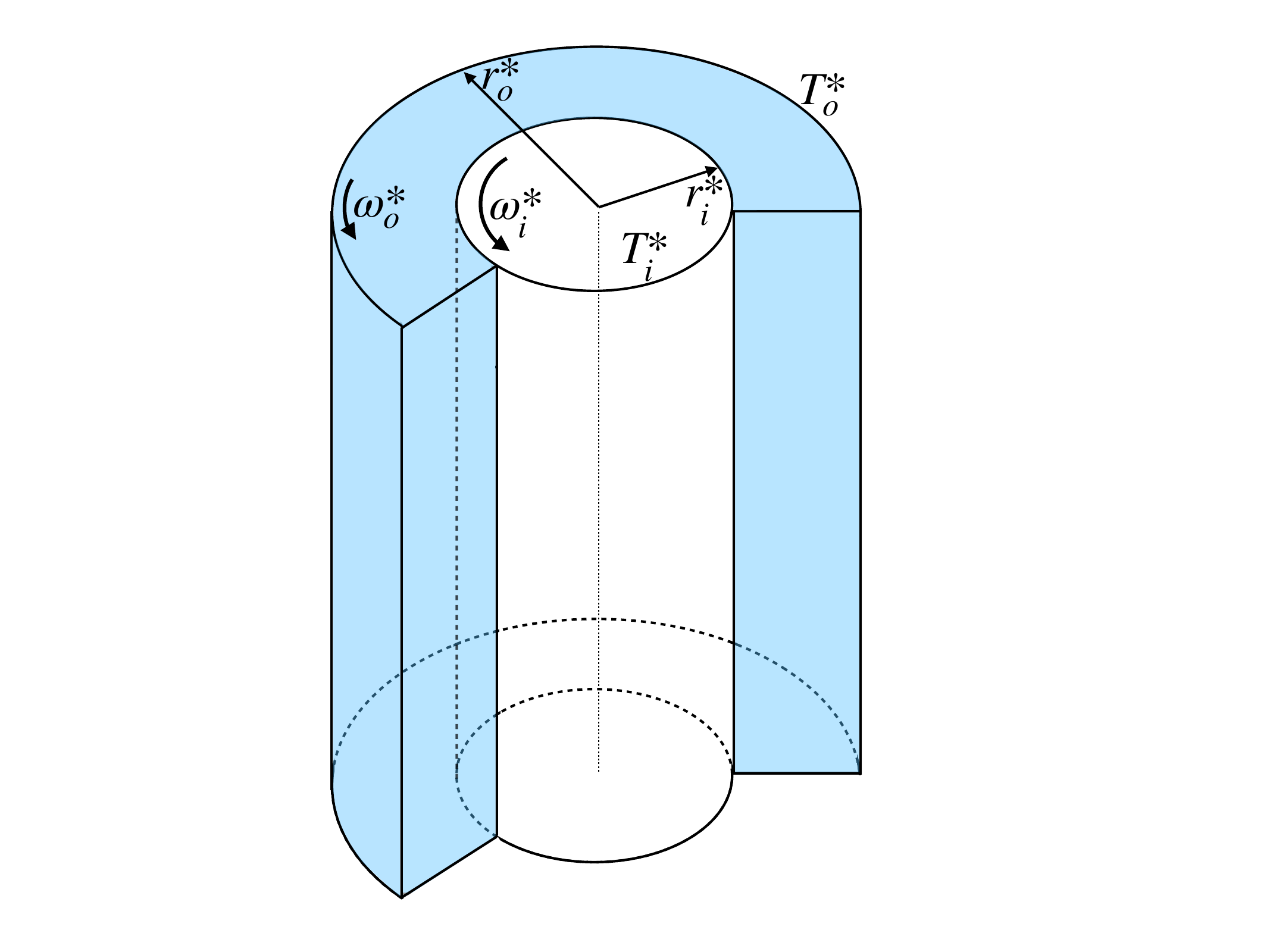}
	\put(-157,150){$(a)$}
	\includegraphics[width=0.45\textwidth,trim= 230 160 230 100,clip]{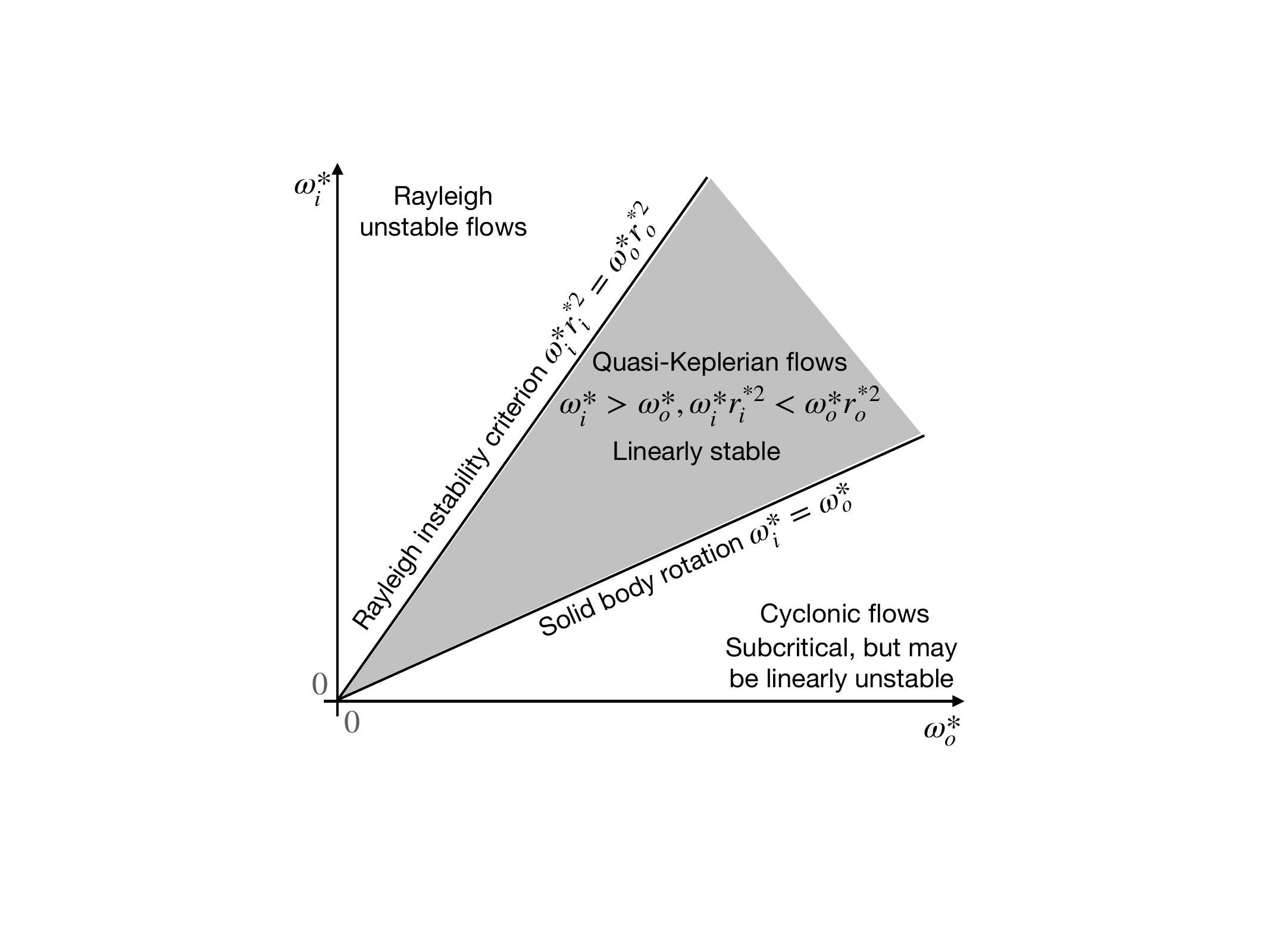}
	\put(-202,150){$(b)$}}	
	\caption{$(a)$ Sketch of the coaxial co-rotating Taylor-Couette flow system investigated in the present work. The gravitational acceleration points radially inwards. $(b)$ Flow regimes in the parameter space. Shaded area corresponds to quasi-Keplerian flows which include the Keplerian case satisfying $\omega_i^* r_i^{*{3}/{2}}=\omega_o^* r_o^{*{3}/{2}}$. It should be noted that the Keplerian flow profile can only be approximated but not exactly satisfied by the radial structure of a TC flow even in the Keplerian case. Thus, the flow at $\omega_i^* r_i^{*{3}/{2}}=\omega_o^* r_o^{*{3}/{2}}$ is also quasi-Keplerian. }
	\label{Fig:schematic_flowregime}
\end{figure}

The flow investigated is an incompressible TC flow confined between two coaxial co-rotating cylinders, differentially heated at the walls, as depicted in figure \ref{Fig:schematic_flowregime}$(a)$. The fluid is characterised by density $\rho^*$, dynamic viscosity coefficient $\mu^*$ (or kinematic viscosity $\nu^*=\mu^*/\rho^*$), and thermal diffusivity coefficient $\kappa^*$. Throughout this paper, dimensional quantities are indicated with the superscript ${}^*$.
For the mathematical modelling of the TC structure, a cylindrical coordinate system $(r,\theta,z)$ is adopted, where $r$, $\theta$, and $z$ represent the radial, azimuthal, and axial directions, respectively. The radius of the inner cylinder is $r_i^*$, and that of the outer cylinder is $r_o^*$. The cylinders are rotating about the $z$ axis in the same direction with angular frequencies $\omega_i^*$ and $\omega_o^*$, respectively. The flow in the $z$ direction is assumed to be homogeneous and infinitely long, allowing the use of periodic boundary conditions in our linear stability analysis.

The wall temperature of the two cylinders is maintained at fixed values: $T_i^*$ on the inner wall and $T_o^*$ on the outer wall, with $T_i^* > T_o^*$, to mimic the temperature distribution in accretion disks \citep{Chiang1997Spectral}. To account for the thermally-driven buoyancy effect due to thermal stratification, stellar gravity in the radial direction is incorporated into the modelling process. Following \cite{Balbus1991Powerful} and \cite{Latter2016Convective}, we adopt the Boussinesq approximation, assuming that the fluid density varies linearly with temperature as $\rho^* = \rho_o^* - \rho_o^* \alpha^* (T^* - T_o^*)$ and this variation is significant only in the gravitational buoyancy term. Here, $\alpha^*$ is the coefficient of thermal expansion, and $\rho_o^*$ is the reference density at temperature $T_o^*$. In contrast to \cite{Meyer2021Stability}, centrifugal buoyancy is ignored in our work; the effect would be stabilising in our temperature setting.

The governing equations consist of the Navier-Stokes equation under the Boussinesq approximation, the continuity equation, and the energy equation expressed in terms of temperature. For non-dimensionalisation, the gap between the two cylinders $d^*=r_o^* - r_i^*$ is chosen as the characteristic length scale, the characteristic velocity scale is the tangential speed of the inner wall $U^*=|\omega_i^* - \omega_o^*|r_i^*$ observed in the rotating frame of reference, the time scale is $d^*/U^*$, and the reference density, pressure and gravitational acceleration magnitude are $\rho_o^*$, $\rho_o^* U^{*2}$, and $g_o^*$, respectively.

The resultant nondimensional equations, expressed in a reference frame rotating with the outer cylinder, are given as follows:
\begin{subequations}\label{eq2.1}
\begin{equation}\label{eq2.1b}
	\frac{\partial \boldsymbol{u}}{\partial t} + \boldsymbol{u} \cdot\boldsymbol{\nabla} \boldsymbol{u} = - \boldsymbol{\nabla}p + \frac{f(\eta)}{{Ta}^{1/2}} \nabla^2 \boldsymbol{u} + \frac{Ra f^2(\eta)}{Pr Ta}T g(r) \boldsymbol{e}_r - \frac{1}{Ro} \boldsymbol{e}_z \times \boldsymbol{u},
\end{equation}
\begin{equation}
\boldsymbol{\nabla} \cdot \boldsymbol{u} = 0,
\end{equation}
\begin{equation}
	\frac{\partial T}{\partial t} + \boldsymbol{u} \cdot\boldsymbol{\nabla} T = \frac{f(\eta)}{Pr {Ta}^{1/2}} \nabla^2 T,
\end{equation}
\end{subequations}
where $f(\eta)={(1+\eta)^3}/{8\eta^2}$. The form of $f(\eta)$ and the introduction of $Ta$ and $Ro$ (defined below) follow the derivations for isothermal flows in TC geometry in \cite{Grossmann2016High}. Here, $\boldsymbol{u}=(u_r,u_\theta,u_z)$ is the non-dimensional velocity vector, $p$ is pressure and $T=(T^* - T_o^*)/(T_i^* - T_o^*)$ is the temperature difference. The associated boundary conditions include the no-slip and no-penetration boundary conditions for the velocity and the Dirichlet  boundary conditions for the temperature, i.e.,
\begin{subequations}
\begin{equation}\label{eq:velo_BC}
	\boldsymbol{u}(r_i) = \boldsymbol{e}_{\theta} \, \text{when $\omega_i^*>\omega_o^*$}, \, \boldsymbol{u}(r_i) = -\boldsymbol{e}_{\theta} \, \text{when $\omega_i^*<\omega_o^*$}, \, \boldsymbol{u}(r_o) = \boldsymbol{0},
\end{equation}
\begin{equation}\label{eq:ri_ro}
	T(r_i) = 1, \ \ \  T(r_o) = 0 \, \, \text{with} \, r_i=\frac{\eta}{1-\eta}, r_o=r_i+1.
\end{equation}
\end{subequations}
Here, $\boldsymbol{e}_{r},\boldsymbol{e}_{\theta},\boldsymbol{e}_{z}$ are the unit vectors in the cylindrical coordinate.

Regarding the gravitational acceleration profile in Eq. \ref{eq2.1b}, it should be noted that from Newton's Law of Universal Gravitation, for uniform mass distribution along the $z$-axis in cylindrical coordinates, the gravitational acceleration magnitude is $g^*(r^*)=2G^* \lambda^* /r^*$, where $G^*$ is the universal gravitational constant and $\lambda^*$ is the mass per unit length of the $z$-axis. After non-dimensionalisation with $g^*(r_o^*)$, it gives the following expression in the non-dimensional form
\begin{equation}\label{eq:gravity}
g(r) = \frac{r_o}{r}.
\end{equation}
To compare, the gravitational potential due to a point mass at the center of the central object in an accretion disk is $\phi^*(r^*)=-G^* M^*/r^{*}$, with the corresponding gravitational acceleration magnitude being $g^*(r^*)=G^* M^*/r^{*2}$ when described in spherical coordinates. Non-dimensionalisation of this $g^*(r^*)$ with $g^*(r_o^*)$ results in a profile of the form $g(r)=r_o^2/r^2$. The third profile that can be considered is an artificial constant profile $g(r)=1$, which has been shown to be experimentally realisable on thermo-electric convection in a cylindrical annulus during a sounding rocket flight \citep{Antoine2023Thermo}. In the result section \ref{sec:results}, the profile given in Eq. \ref{eq:gravity} is studied. As suggested by one of the reviewers, a brief comparison of the effects of the other two profiles with Eq. \ref{eq:gravity} is conducted; see section 1 of the supplementary material.

In this equation system, there are five governing parameters defined as
\begin{equation}
\eta = \frac{r_i^*}{r_o^*},\ \ \ Ta = \frac{(1+\eta)^4}{64 \eta^2}\frac{d^{*2}(r_i^* + r_o^*)^2 (\omega_i^* - \omega_o^*)^2}{\nu^{*2}},\ \ \ Pr = \frac{\nu^*}{\kappa^*},\ \ \ Ra = \frac{\alpha^* g_o^*(T_i^* - T_o^*)d^{*3}}{\nu^* \kappa^*}, \notag
\end{equation}
\begin{equation}\label{eq:five_parameters}
Ro=\frac{|\omega_i^* - \omega_o^*|r_i^*}{2\omega_o^* d^*} = \frac{\eta |1-a|}{2(1-\eta)a} \ \text{with the rotation ratio} \ a=\frac{\omega_o^*}{\omega_i^*}=\eta^{q},
\end{equation}
where the rotation exponent parameter $q$ is real-valued for co-rotating cylinders ($a>0$), the case considered in the present study. Specifically, when $\omega_i^*>\omega_o^*>0$, we have $0<a<1$ and thus $q>0$.
\begin{itemize}
\item Among these parameters, the radius ratio $\eta$ determines the cylinders' gap width as well as inner and outer radii $r_i$ and $r_o$ since we require $r_o-r_i=1$ according to the non-dimensionalisation step. Small values of $\eta$ correspond to large accretion disks with small central stars.
\item The Taylor number $Ta$ defines the relative importance of fluid inertia relative to viscosity. For highly rarefied gases in fast rotating motions, $Ta$ can be extremely high, which will be considered in our work.
\item The Prandtl number $Pr$ for most gases is of order one, meaning that momentum and heat diffuse at comparable rates. For rarefied gases, $Pr$ would be smaller.
\item The Rayleigh number $Ra$ characterises the importance of natural thermal convection relative to thermal diffusion in the quasi-Keplerian flow. Weak thermal stratification dictates small values of $Ra$.
In addition to $Ra$, we will also adopt the Richardson number $Ri\equiv Ra/(PrTa)$ which characterises the relative importance between the convection due to thermal buoyancy and the flow shearing. 
\item Finally the Rossby number $Ro$ quantifies the relative rotation rate of the two cylinders. It should be emphasised that for a given $\eta$, $Ro$ depends only on $q$. The Keplerian regime corresponds to a specific rotation scenario as discussed below.
\end{itemize}

The rotation exponent $q$ is important in delimiting the various flow regimes illustrated in figure \ref{Fig:schematic_flowregime}$(b)$: $q=0$ indicates solid-body rotation; $q>2$ corresponds to the conventional Rayleigh unstable regime of isothermal flows; $0<q<2$, resulting in $\omega_i^* r_i^{*2} < \omega_o^* r_o^{*2}$, represents the Rayleigh stable regime where the angular momentum increases monotonically with $r$. The special case at $q=3/2$, where the rotation rates of the two cylinders obey Kepler's third law, is referred to as the Keplerian regime. Conventionally, the entire range of $0<q<2$ is referred to as the quasi-Keplerian regime. The present study focuses on the case at $q=3/2$.

\subsection{Linear stability analysis: inhomogeneous in the radial direction}\label{sec:global_stability_analysis}

The equation system \ref{eq2.1}-\ref{eq:five_parameters} described above admits the following one-dimensional laminar steady solution, which can be theoretically derived by noting that the only non-zero velocity component is the azimuthal velocity and the solution is a function of $r$. For the investigated case of $\omega_i^*>\omega_o^*>0$, the laminar solution reads
\begin{equation}
U_{b,r}=0, U_{b,\theta}(r)=\frac{1}{\eta(1+\eta)}\left(\frac{r_i^2}{r} - \eta^2 r\right),U_{b,z}=0,T_b(r) = \frac{1}{\text{ln}\eta}\text{ln}r - \frac{\text{ln}r_o}{\text{ln}\eta}.
\end{equation}
Here the pressure profile of the laminar solution $P_b(r)$ is not provided since it is not needed in the linear stability analysis.

To probe the stability or instability of the above laminar base flow subject to disturbances, we decompose the total flow variables in Eq. \ref{eq2.1} into their base states plus perturbations
\begin{equation}
\boldsymbol{u}=\boldsymbol{U}_b + \boldsymbol{u}', \, \, p = P_b + p', \, \, T = T_b + T'.
\end{equation}
Then linearisation around the base flow leads to the following linear perturbation equation
\begin{subequations}\label{eq:perturbation_phys}
\begin{equation}
	\frac{\partial \boldsymbol{u}'}{\partial t} + \boldsymbol{u}' \cdot\boldsymbol{\nabla} \boldsymbol{U}_b + \boldsymbol{U}_b \cdot\boldsymbol{\nabla} \boldsymbol{u}'= - \boldsymbol{\nabla}p' + \frac{f(\eta)}{{Ta}^{1/2}} \nabla^2 \boldsymbol{u}' + \frac{Ra f^2(\eta)}{Pr Ta}T' g(r) \boldsymbol{e}_r - \frac{1}{Ro} \boldsymbol{e}_z \times \boldsymbol{u}',
\end{equation}
\begin{equation}
\boldsymbol{\nabla} \cdot \boldsymbol{u}' = 0,
\end{equation}
\begin{equation}
	\frac{\partial T'}{\partial t} + \boldsymbol{U}_b \cdot\boldsymbol{\nabla} T' + \boldsymbol{u}' \cdot\boldsymbol{\nabla} T_b = \frac{f(\eta)}{Pr {Ta}^{1/2}} \nabla^2 T',
\end{equation}
\end{subequations}
along with homogeneous Dirichlet boundary conditions for both the perturbed velocity and temperature. The solution to Eq. \ref{eq:perturbation_phys} can be assumed to take the normal-mode form, i.e.,
\begin{equation}\label{eq:wavelike}
(\boldsymbol{u}',p',T') = [\tilde{\boldsymbol{u}}'(r),\tilde{p}'(r),\tilde{T}'(r)]\text{exp}(-i\omega t + im\theta+ik z) + \text{c.c.}, 
\end{equation}
where integer-valued $m$ and real-valued $k$ are the azimuthal and axial wavenumbers, respectively. Variables with the symbol tilde $\tilde{}$ are shape functions in the radial direction and c.c. represents the complex conjugate of the preceding term. Additionally, $\omega=\omega_r + i\omega_i$ includes $\omega_r$ as the frequency and $\omega_i$ as the linear growth rate. At a given parameter setting, the flow is said to be linearly unstable if $\omega_i>0$, linearly stable if $\omega_i<0$ and neutral if $\omega_i=0$.

Inserting Eq. \ref{eq:wavelike} into the perturbation equations \ref{eq:perturbation_phys} results in the linearised equations in the spectral space, which are given as follows
\begin{subequations}\label{eq:perturbation_spec}
\begin{align}\label{eq:perturbation_spec_ur}
-i\omega \tilde{u}'_r = &\left[-\frac{imU_{b,\theta}}{r} + \frac{f(\eta)}{{Ta}^{1/2}} \left(\tilde{\nabla}^2 - \frac{1}{r^2}\right)\right]\tilde{u}'_r \notag\\
&+ \left[\frac{2U_{b,\theta}}{r} + \frac{f(\eta)}{{Ta}^{1/2}} \left(-\frac{2im}{r^2}\right) + \frac{1}{Ro} \right] \tilde{u}'_{\theta}- \frac{d \tilde{p}'}{d r} + \frac{Ra f^2(\eta)}{Pr Ta}\tilde{T}' g(r),
\end{align}
\begin{align}\label{eq:perturbation_spec_ut}
	-i\omega \tilde{u}'_{\theta} = &\left[-\frac{d U_{b,\theta}}{d r}-\frac{U_{b,\theta}}{r} + \frac{f(\eta)}{{Ta}^{1/2}} \left(\frac{2im}{r^2}\right)- \frac{1}{Ro} \right]\tilde{u}'_r \notag\\
	&+ \left[-\frac{imU_{b,\theta}}{r} + \frac{f(\eta)}{{Ta}^{1/2}} \left(\tilde{\nabla}^2 - \frac{1}{r^2}\right)  \right] \tilde{u}'_{\theta}- \frac{im \tilde{p}'}{ r},
\end{align}
\begin{align}
	-i\omega \tilde{u}'_{z} = \left(-\frac{imU_{b,\theta}}{r} + \frac{f(\eta)}{{Ta}^{1/2}} \tilde{\nabla}^2  \right) \tilde{u}'_{z} - ik \tilde{p}',
\end{align}
\begin{align}
0 = \left(\frac{d}{d r} +  \frac{1}{r} \right) \tilde{u}'_{r} +\frac{im }{ r}\tilde{u}'_{\theta} + ik \tilde{u}'_{z},
\end{align}
\begin{align}
	-i\omega \tilde{T}' = -\frac{d T_b}{d r} \tilde{u}'_{r} +  \left(-\frac{imU_{b,\theta}}{r} + \frac{f(\eta)}{Pr {Ta}^{1/2}} \tilde{\nabla}^2 \right)   \tilde{T}',
\end{align}
\end{subequations}
where the Laplacian $\tilde{\nabla}^2 = d^2 /dr^2 + 1/r (d/dr) - m^2/r^2 - k^2$.
The above equation system can be written in a generalised eigenvalue problem in the form of 
\begin{equation}\label{eq:compact_form_eig}
-i\omega \widetilde{\boldsymbol{M}} \tilde{\boldsymbol{\gamma}}=\widetilde{\boldsymbol{L}} \tilde{\boldsymbol{\gamma}},
\end{equation}
or more specifically
\begin{equation}\label{eq:eigenvalue_problem}
-i\omega
\begin{pmatrix}
	1 & 0 & 0 & 0 & 0\\
	0 & 1 & 0 & 0 & 0 \\
	0 & 0 & 1 & 0 & 0 \\
	0 & 0 & 0 & 0 & 0 \\
	0 & 0 & 0 & 0 & 1
\end{pmatrix}
\begin{pmatrix}
	\tilde{u}'_{r} \\ \tilde{u}'_{\theta} \\ \tilde{u}'_{z} \\ \tilde{p}' \\ \tilde{T}'
\end{pmatrix} = 
\begin{pmatrix}
	L_{r,r} & L_{r,\theta} & 0 & L_{r,p} & L_{r,T}\\
	L_{\theta,r} & L_{\theta,\theta} & 0 & L_{\theta,p} & 0 \\
	0 & 0 & L_{z,z} & L_{z,p} & 0 \\
	L_{p,r} & L_{p,\theta} & L_{p,z} & 0  & 0 \\
    L_{T,r} & 0 & 0 & 0 & L_{T,T}
\end{pmatrix}
\begin{pmatrix}
	\tilde{u}'_{r} \\ \tilde{u}'_{\theta} \\ \tilde{u}'_{z} \\ \tilde{p}' \\ \tilde{T}'
\end{pmatrix}.
\end{equation}
Here the linear operators $L_{*,*}$ can be derived by matching the equations term by term. The perturbation variable vector $\tilde{\boldsymbol{\gamma}}=(\tilde{u}'_{r},\tilde{u}'_{\theta},\tilde{u}'_{z},\tilde{p}',\tilde{T}')^T$ is identified as the eigenvector. Since the eigenvector can be arbitrarily scaled in the linear analysis, to eliminate ambiguity, it is normalised to have a unit amplitude and zero phase angle in the azimuthal velocity component at the middle point of the gap $\tilde{u}'_{\theta}(r={(r_i+r_o)}/{2})$, unless specified otherwise.

\subsection{Linear stability analysis: local (or homogeneous) in the radial direction}

Local stability analysis is a widely used method for investigating flow instability in modelled accretion disk problems \citep{Klahr2014Convective,Lyra2014Convective,Latter2016Convective,Held2018Hydrodynamic}. 
Among these studies,  \cite{Klahr2014Convective} and \cite{Lyra2014Convective} employed a short wave approximation in their local analysis to explore the convective overstability in radially stratified disks under thermal relaxation. In contrast,  \cite{Latter2016Convective} and \cite{Held2018Hydrodynamic} utilised the shearing box approximation to analyse vertically stratified disk models. Following the suggestion from one of the reviewers, we conduct a local stability analysis of the problem formulated in section \ref{sec:global_stability_analysis} and compare the results with those from our global analysis (global in the radial direction, yet local in the axial direction).

Given that the flow under investigation is radially stratified, the shearing box approximation in a local rotating Cartesian coordinate system is unsuitable for formulating an eigenvalue problem. The main challenge is implementing the shearing periodic boundary condition in the radial direction along which thermal stratification and gravitational acceleration exist. While the implementation is feasible for numerical simulations, it is not appropriate for the present analysis. Therefore, we adopt the short wave approximation, following the approach of \cite{Klahr2014Convective} and \cite{Lyra2014Convective}. In this approximation, perturbations are expressed in the Fourier mode form as follows
\begin{equation}\label{eq:shortwave}
(\boldsymbol{u}',p',T') = [\hat{\boldsymbol{u}}',\hat{p}',\hat{T}']\text{exp}(-i\omega t + i k_r r + im\theta+ik_z z) + \text{c.c.}. 
\end{equation}
Note the distinction from Eq. \ref{eq:wavelike}: wavelike solutions are also assumed in the radial direction here, rendering this analysis fully local in all three directions. In this context, $k_r$ and $k_z$ represent the radial wavenumber and axial wavenumber, respectively. Noting that the linear instability to be reported in section \ref{sec:results} below is predominantly axisymmetric ($m=0$), in this local analysis we focus on the case where $m=0$, meeting the assumptions $m\ll k_r r$ and $m\ll k_z z$ for the short wave approximation.

Inserting Eq. \ref{eq:shortwave} into the perturbation equation \ref{eq:perturbation_phys} leads to the following equation system for the local quasi-Keplerian flow at a radial location $r_0$
\begin{subequations}\label{eq:perturbation_spec_shortwave}
	\begin{equation}
	-i\omega \hat{u}'_r =  \frac{f(\eta)}{{Ta}^{1/2}} \hat{\nabla}^2\hat{u}'_r + \left(\frac{2U_{b,\theta}(r_0)}{r_0}  + \frac{1}{Ro} \right) \hat{u}'_{\theta}- i k_r \hat{p}' + \frac{Ra f^2(\eta)}{Pr Ta}\hat{T}' g(r_0),
	\end{equation}
	\begin{equation}
	-i\omega \hat{u}'_{\theta} = \left[-\left(\frac{d U_{b,\theta}}{d r}\right)_{r_0}-\frac{U_{b,\theta}(r_0)}{r_0} - \frac{1}{Ro} \right]\hat{u}'_r + \frac{f(\eta)}{{Ta}^{1/2}} \hat{\nabla}^2 \hat{u}'_{\theta},
	\end{equation}
	\begin{equation}
	-i\omega \hat{u}'_{z} =\frac{f(\eta)}{{Ta}^{1/2}} \hat{\nabla}^2 \hat{u}'_{z} - ik_z \hat{p}',
	\end{equation}
	\begin{equation}
	0 = i k_r \hat{u}'_{r} + ik_z \hat{u}'_{z},
	\end{equation}
	\begin{equation}
	-i\omega \hat{T}' = -\left(\frac{d T_b}{d r}\right)_{r_0} \hat{u}'_{r} +  \frac{f(\eta)}{Pr {Ta}^{1/2}} \hat{\nabla}^2   \hat{T}',
	\end{equation}
\end{subequations}
where the Laplacian $\hat{\nabla}^2 = - k_r^2 - k_z^2$. In this analysis, some less significant terms have been omitted, following \cite{Lyra2014Convective}. For instance, for an infinitesimal disturbance to the local flow at $r_0$, it is assumed that $u'_r/r\ll \partial_r u'_r$. Consequently, the term $u'_r/r$ in the continuity equation is neglected. Similarly, we neglect the term $1/r\partial _r u'_r$ compared to $\partial_{rr} u'_r$ in the Laplacian. The resulting equations \ref{eq:perturbation_spec_shortwave} form an $5\times 5$ eigenvalue problem, which is computationally much more efficient to solve than Eq. \ref{eq:eigenvalue_problem}. Note that no discretisation is needed for solving Eq. \ref{eq:perturbation_spec_shortwave} in the local analysis. In contrast, Eq. \ref{eq:eigenvalue_problem}, after discretisation based on the spectral collocation method, results in a $5N\times 5N$ eigenvalue problem for the global analysis presented in section \ref{sec:results}, where the node number $N$ ranges from 51 for normal parameters to 601 for extreme parameters in our calculations.

\section{Numerical method and validation}\label{sec:methods_validations}

To solve the generalised linear eigenvalue problem expressed in Eq. \ref{eq:compact_form_eig}, we developed an in-house MATLAB code using the spectral collocation method \citep{Trefethen2000Spectral}. This approach employs Chebyshev-Lobatto nodes, clustered near both the inner and outer cylinders while being sparse at the gap center. For Chebyshev polynomials of $(N-2)$th order, there are in total $N$ Chebyshev-Lobatto nodes. In our calculations, a resolution of $N=51$ typically proves sufficient, although for extreme parameter regimes, we extend up to $N=601$. Central to our methodology is the construction and utilisation of Chebyshev differentiation matrices, facilitating the formation of $\widetilde{\boldsymbol{M}}$ and $\widetilde{\boldsymbol{L}}$, as described in \cite{Trefethen2000Spectral}. Implementing homogeneous boundary conditions for $\tilde{\boldsymbol{u}}'$ and $\tilde{T}'$ at the inner cylinder $r=r_i$ and outer cylinder $r=r_o$ is straightforward, accomplished by removing the corresponding rows and columns from $\widetilde{\boldsymbol{M}}$ and $\widetilde{\boldsymbol{L}}$. Subsequently, we employ MATLAB's built-in functions `eig()' and `eigs()' to solve the generalised linear eigenvalue problem \ref{eq:compact_form_eig}, yielding the eigenvalue $\omega$ and the eigenvector $\tilde{\boldsymbol{\gamma}}$. A validation against the data in \cite{Deguchi2017Linear} and a convergence test of our code can be found in section 2 of the supplementary material; favourable agreement and good convergence have been achieved.
For the local analysis, the small eigenvalue problem \ref{eq:perturbation_spec_shortwave} is also solved using the two built-in functions.

\section{Results and discussion}\label{sec:results}

To clearly present the numerical results below, a summary of the following subsections is provided at the outset.
\begin{itemize}
\item Subsection \ref{firstsection} characterises the thermo-hydrodynamic linear instability at a moderate Taylor number of $Ta=10^6$ in the TC flow. While this value may not reach the extreme parameters relevant to accretion disk dynamics, its selection facilitates a clear elucidation of the primary destabilising mechanism. We will demonstrate the robustness of the flow pattern in the corresponding unstable mode, noting that even at significantly higher $Ta$, the eigenvector pattern bears resemblance to that observed at $Ta=10^6$. 
\item Subsection \ref{sec:effects_of_parameters} delves into an exploration of the effects of the principal governing parameters, including $Ta$, $Ri$, $Pr$ and $\eta$. In a linear stability analysis, a key point of interest is to determine the lowest Taylor number, referred to as the critical Taylor number $Ta_c$, beyond which the linear instability sets in. Notably, we illustrate that the instability exhibits resilience even at very low yet finite Richardson numbers $Ri\sim 0.01$, indicative of weak thermal effects. The critical $Ta_c$ appears to scale with the deviation of $Ri$ from the limiting $Ri$ corresponding to infinite $Ta_c$. Additionally, another scaling law emerges at asymptotically small Prandtl numbers $Pr\sim 10^{-4}$. By extrapolating these findings to high $Ta$ regimes, we infer the pertinence of our observations to the dynamics of accretion disks. Particularly, section \ref{app:comparison_local} is the only part dedicated to the local stability analysis for comparison with the global one. Qualitatively consistent results is obtained; differences are underscored.
\item The final subsection \ref{sec:extreme_case} is dedicated to an examination of an extreme case scenario characterised by $\eta=0.05$, $Pr=10^{-3}$, $Ri=10^{-3}$, and $10^{9}\lessapprox Ta \lessapprox 10^{16}$. Within the linear framework, this scenario mimics the hydrodynamics in accretion disks based on the TC flow geometry.
\end{itemize} 

\subsection{Characteristics of the thermo-hydrodynamic linear instability}\label{firstsection}

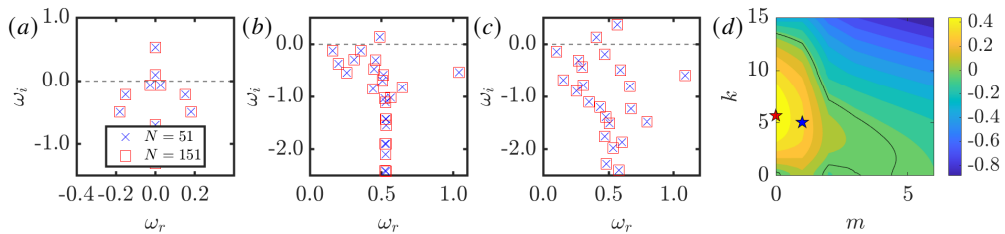
\begin{figure}
	\centering
	\includegraphics[width=0.99\textwidth,trim= 0 0 0 0,clip]{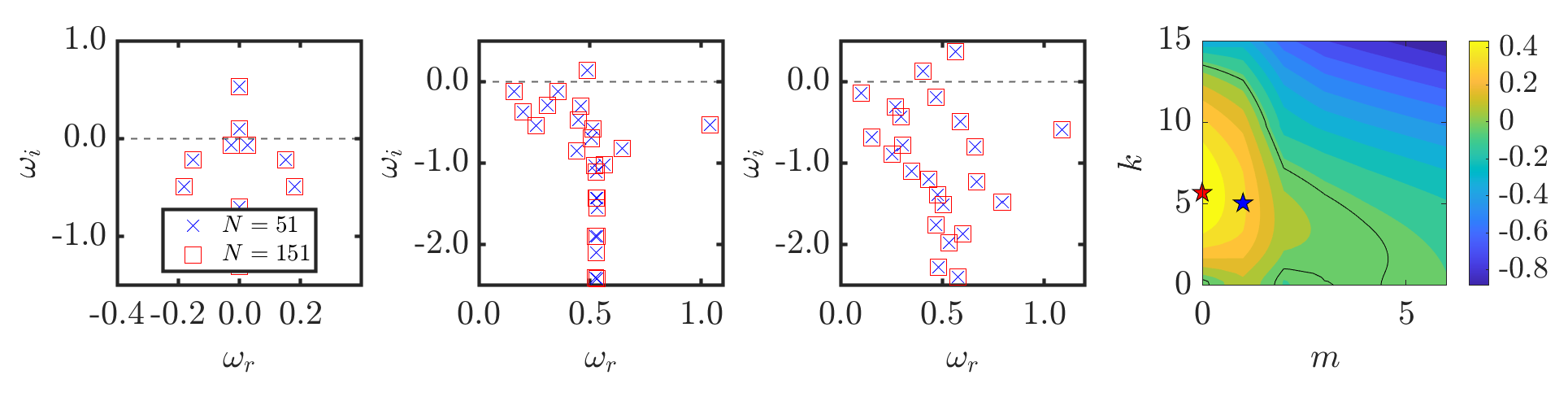}
	\put(-378,80){$(a)$}
	\put(-290,80){$(b)$}
	\put(-203,80){$(c)$}
	\put(-113,80){$(d)$}			
	\caption{Eigenspectra of the flow in the Keplerian regime at $\eta=0.3$, $Pr=0.7$, $Ri=0.1$ and $Ta={10}^6$ for: $(a)$ $(k,m)=(5,0)$; $(b)$ $(k,m)=(0,1)$; $(c)$ $(k,m)=(5,1)$. The modes above the dashed line are linearly unstable. $(d)$ The corresponding linear growth rate $\omega_i$ in the $(k,m)$ plane where the black curve traces $\omega_i=0$. Note that $m$ is a non-negative integer; at each combination of $(k,m)$, $\omega_i$ of the most unstable/least stable mode in an eigenspectrum is recorded for the plot. The red and blue stars in panel $(d)$ mark respectively the locations of the most unstable mode for all $(k,m)$ and the most unstable non-axisymmetric mode for all $k$ and $m\neq0$. The corresponding mode patterns are visualised in figure \ref{Fig:axisymmetric_mode}. }
	\label{Fig:eigen_convergence}
\end{figure}

Within the Keplerian regime, characterised by $q=1.5$, we observe a linear instability within the thermal TC flow. The dynamics of this instability exhibit remarkably rich complexity. To illustrate this, we maintain fixed values for the governing parameters: $\eta=0.3$, $Pr=0.7$, $Ri=0.1$ and $Ta={10}^6$.
The selection of these parameters is guided by several considerations. Firstly, the chosen radius ratio, though relatively small, remains achievable in both experimental setups and numerical simulations. This paves the way for investigating flow instability in future works, particularly those exploring its nonlinear evolution and potential turbulence simulations.
Secondly, the specific $Pr$ chosen reflects a typical value for most gases under atmospheric conditions. However, it is important to note that while hydrogen and helium primarily constitute the gas components in accretion disks \citep{Pringle1981Accretion,Mineshige1993Accretion,Ikoma2012Situ,Klahr2014Convective}, the corresponding $Pr$ therein can be significantly smaller \citep{Latter2016Convective}. We will investigate such extreme cases in subsections \ref{sec:effects_of_parameters} and \ref{sec:extreme_case}. Lastly, the selected $Ri$ aligns with commonly used values in the literature \citep{Lesur2010Angular,Held2018Hydrodynamic}. The variation of these parameters will be explored in the next subsection.

Figure \ref{Fig:eigen_convergence}$(a)$ displays the converged results with $N=(51,151)$ for the eigenspectra of the axisymmetric ($m=0$) modes, where the two modes above $\omega_i=0$ (the dashed line) are linearly unstable. These are stationary eigenmodes with zero real parts, meaning that they grow over time but do not travel in space. 
Non-axisymmetric modes ($m\ne0$) can also become linearly unstable, further classified into two types. One type is axial-independent ($k=0$) with a single unstable mode, as shown in panel \ref{Fig:eigen_convergence}$(b)$, and the other is axial-dependent ($k\neq 0$) with two unstable modes, also called helical modes, as displayed in panel \ref{Fig:eigen_convergence}$(c)$. The frequencies of these unstable non-axisymmetric modes are non-zero, meaning they oscillate in time and travel in the azimuthal direction and also in the axial direction if $k\neq0$.

In the above, we have reported two categories of unstable modes: stationary axisymmetric modes and oscillatory non-axisymmetric modes. However, axisymmetric modes can also become oscillatory at specific parameters, as will be shown in figure \ref{Fig:eta005_Ta1e16_m01}$(b)$.
In a different TC flow system with a geophysical application background, \cite{Jenny2007Primary} reported the coexistence of these three kinds of mode. The difference between their study and ours is that they fixed the outer cylinder and examined the radial buoyancy effects due to temperature and/or salinity stratification in the Earth's equatorial region. The difference between the two settings, along with the similarity in the results, demonstrates the robustness of the instability in thermal TC flows with different configurations. Distinguishing these different unstable modes is not only of interest within the linear framework but also crucial from a (weakly) nonlinear perspective when examining flow bifurcations. For example, \cite{Kang2017Radial} studied the effect of centrifugal buoyancy on circular Couette flow and found that the flow bifurcation to stationary axisymmetric modes was always supercritical, while that to oscillatory axisymmetric modes was subcritical. Therefore, it would also be interesting to examine the weak nonlinearity of the present thermally-stratified quasi-Keplerian flow in future studies.

\begin{figure}
	\centering
	\includegraphics[width=0.49\textwidth,trim= 0 0 0 0,clip]{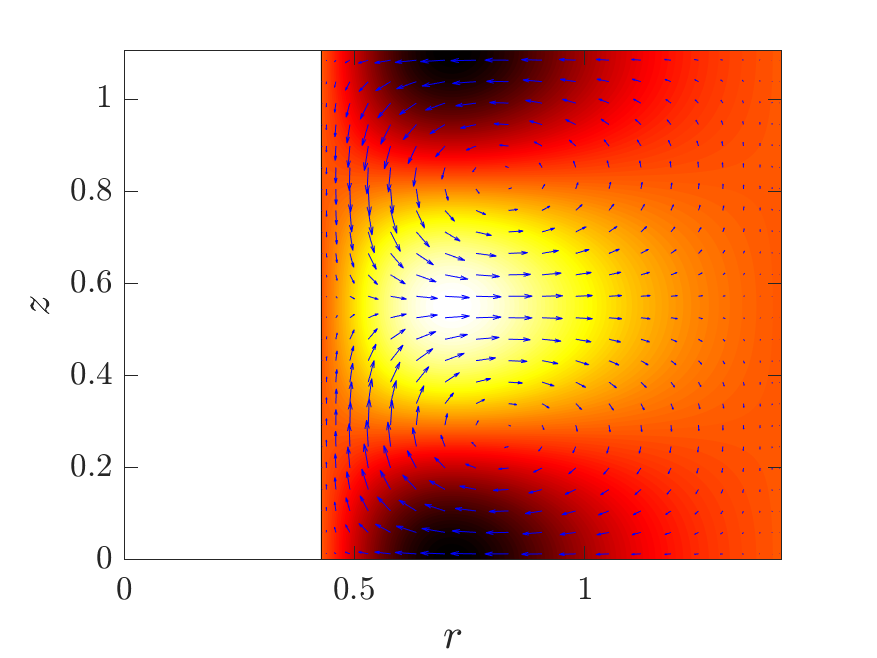}
	\put(-185,120){$(a)$}
	\includegraphics[width=0.49\textwidth,trim= 0 0 0 0,clip]{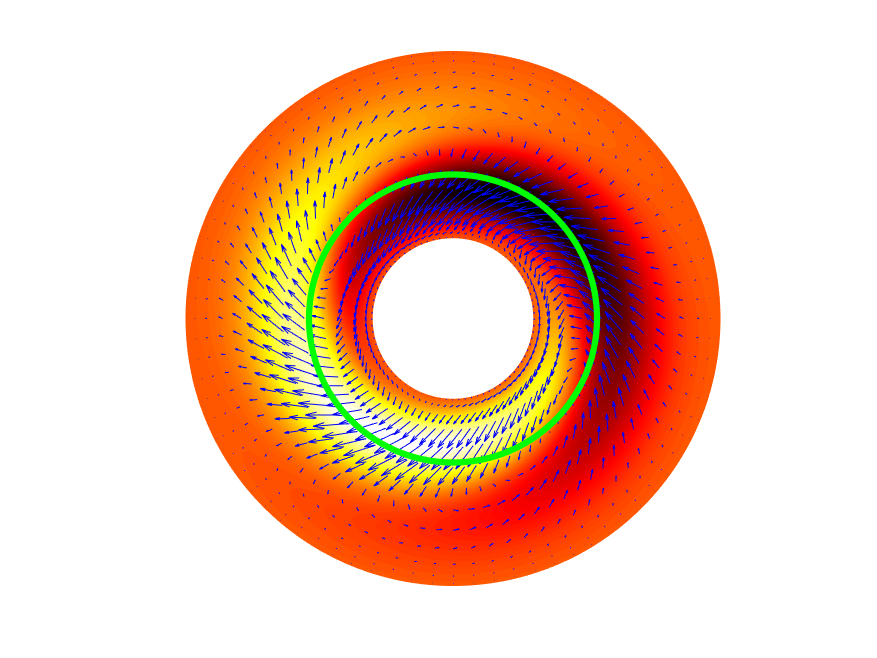}
	\put(-145,120){$(b)$}
	\caption{Contours of disturbance temperature in the quasi-Keplerian flow at $\eta=0.3$, $Pr=0.7$, $Ri=0.1$ and $Ta={10}^6$. Panel $(a)$ is for the most unstable mode attained at $k\approx 5.685$ and $m=0$, corresponding to the red star marked in figure \ref{Fig:eigen_convergence}$(d)$. Panel $(b)$ is for the leading non-axisymmetric mode at $k\approx 5.049$ and $m=1$, corresponding to the blue star marked in figure \ref{Fig:eigen_convergence}$(d)$. In panel $(a)$ one wavelength $2\pi/k\approx 1.11$, comparable to the cylinder gap $d=1$, is shown in the $z$ direction. The green circle in panel $(b)$ marks the location of the critical layer. }
	\label{Fig:axisymmetric_mode}
\end{figure}

To gain a comprehensive view of the most unstable/least stable modes in the axial- and azimuthal-wavenumber space, we plot the linear growth rate $\omega_i$ in the $k$-$m$ plane in panel \ref{Fig:eigen_convergence}$(d)$. The largest growth rate $\omega_i \approx 0.5332736016$, is observed at $m=0$ and $k\approx 5.685$ (red star in panel $d$). This indicates that the most unstable mode in the quasi-Keplerian flow at $(q,\eta,Pr,Ri,Ta)=(1.5,0.3,0.7,0.1,10^6)$ is axisymmetric.
To examine the pattern of this mode, we visualise its temperature contours superposed with the velocity field in the $r$-$z$ plane in figure \ref{Fig:axisymmetric_mode}$(a)$. From this figure, we observe that the axial scale of the mode is comparable to the gap width, with the velocity field representing a pair of counter-rotating vortices aligned in the periodic $z$ direction.
This convection feature is reminiscent of the canonical Rayleigh-B{\'e}nard convection \citep{Bodenschatz2000Recent} where gravitational buoyancy drives fluid from the hot wall to the cold wall, resulting in two counter-rotating vortices per wavelength. This suggests that the thermal effects dominate in rendering the instability in quasi-Keplerian flows, as the Keplerian shearing is known to be strongly stabilising.

In addition to the above most unstable axisymmetric mode, the helical mode with the largest linear growth rate $\omega_i=0.3658437831$, attained at $m=1$ and $k\approx 5.049$ (blue star in figure \ref{Fig:eigen_convergence}$d$), is visualised in figure \ref{Fig:axisymmetric_mode}$(b)$ in the $r$-$\theta$ plane at $z=0$. This mode manifests a spiral structure rotating counterclockwise along with the laminar base flow.  The two ends of the two spiral arms are located where hot fluids decelerate upon encountering the outer wall and then turn back towards the inner wall. Scrutinising the velocity vector field reveals that the mode structure actually consists of two counter-rotating vortices severely distorted by the Keplerian shearing in the bulk flow.
The frequency of the above most unstable helical mode is $\omega_r=0.5662622568$, corresponding to a phase speed $c_r = \omega_r / m\approx0.566$. This indicates that the disturbance travels in the same direction as the bulk flow, with a speed about $24\%$ of the maximum angular frequency $2.333$ at the inner cylinder wall. Since the angular frequency profile of the laminar base flow can be calculated as $\Omega=U_{b,\theta}/r$, with $\eta=0.3$, $r_o=1/(1-\eta)\approx 1.429$, and $r_i=r_o - 1\approx 0.429$, the above phase speed $c_r$ is equal to the local value of $\Omega$ at a location $r\approx 0.769$. This position, where the azimuthal phase speed of a mode is locally equal to the angular velocity of the base flow, is identified as the critical layer in curvilinear shear flows, a concept which has been employed by \cite{Leclercq2016Nonlinear} in their study on stratified TC flow. In the present study, this position, marked by a green circle in figure \ref{Fig:axisymmetric_mode}$(b)$, is located $0.66$ length-scale from the outer wall and $0.34$ from the inner wall. From the perspective of the critical-layer concept developed in the context of parallel shear flows, this location is identified as the most receptive position for energy amplification \citep{Maslowe1986,McKeon2017}. It can be seen that the critical layer in figure \ref{Fig:axisymmetric_mode}$(b)$ approximately coincides with the radial location where the disturbance magnitude is the largest, suggesting its relevance in the present curvilinear context. Nevertheless, its importance remains to be examined in the future.

Although the above results pertain to a moderate Taylor number ($Ta=10^6$), which is not high enough to be directly linked to accretion disks, the mode patterns and underlying instability mechanisms can be applied to high $Ta$ regimes. Numerical evidence at $Ta=(10^8, 10^{10})$ is provided in section 4 of the supplementary material. We have checked that the patterns persist even at $Ta=10^{16}$, but the axial length scale is too small and thus not shown. As $Ta$ increases, both the most unstable axisymmetric mode and the helical mode become increasingly localised around the inner rotating cylinder, suggesting that instability space is diminished by the enhanced Keplerian shearing from fast rotation.
This localisation can be relevant for the accretion-disk dynamics for two reasons. First, accretion disks are typically flat and thin structures \citep{Abramowicz2014Accretion}, making it difficult for disturbances with large axial wavelengths to survive, leaving short-wavelength disturbances as primary candidates for triggering flow transitions. For example, \cite{Klahr2014Convective} adopted a quasi-hydrostatic approximation in their study of convective overstability in radially stratified accretion disks, requiring perturbations to be vertically thin. Additionally, \cite{Latter2016Convective} estimated that the dominant vertical wavelength of flow structures in a protoplanetary disk is about $1\%$ of the disk  thickness at 1 AU. Second, \cite{Latter2016Convective} found that the convective overstability discovered by \cite{Klahr2014Convective} was not prevalent in the outer disk regions but seemed only possible in the inner regions, which may be corresponding to the narrow region around the inner rotating cylinder at high $Ta$ in our study.
Comparatively, it is interesting to note that for vertical thermal stratification under vertical stellar gravity, \cite{Held2018Hydrodynamic} found that the fastest growing mode was at the shortest radial length scales, manifesting as radially thin elongated structures.

From these perspectives, our findings are generally consistent with previous research on thermally-driven linear instability in accretion disks. However, previous studies often used 2-D local shearing rectangles \citep{Klahr2014Convective,Latter2016Convective} or 3-D local shearing boxes \citep{Lyra2014Convective,Held2018Hydrodynamic}, assuming homogeneity in all directions of a cylindrical coordinate system. These studies bear local significance, whereas the linear instability in our work is of a 3-D global nature in the TC configuration. Clearly, the mode structures observed in our study cannot be fully contained within their local shearing boxes.
As suggested by one of the reviewers, to evaluate the similarities and differences between the present global analysis and the local analysis of the same quasi-Keplerian flow, we have also conducted a local analysis based on the short wave approximations employed in \cite{Klahr2014Convective} and \cite{Lyra2014Convective}. The results, as will be presented in section \ref{app:comparison_local}, are generally consistent with those from the global analysis. However, quantitatively, some scaling laws (to be discussed in the next subsection) differ. In addition, none of the unstable mode structures observed in the global analysis can be captured in the local analysis.
Besides, the relevant work \cite{Meyer2021Stability} also did not arrive at the same conclusion presented above as they focused on small $Ta$ in their stability analysis.

To sum up, the analysis in this subsection indicates that the destabilising thermally-driven buoyancy can overcome the stabilising Keplerian shear to render the linear instability in the quasi-Keplerian flow.
As will be further demonstrated in section \ref{sec:extreme_case}, this linear instability persists even when the thermal buoyancy effect is very weak but finite, despite the strong Keplerian shear. It can arise at exceptionally large axial wavenumbers (small axial wavelengths), which are especially relevant to the flat geometry of accretion disks. This demonstrates the potential prevalence of thermo-hydrodynamic linear instability in accretion disks, particularly in the inner radii regions, which is one of the most significant findings of this study.

\subsection{Effects of governing parameters on the flow instability}\label{sec:effects_of_parameters}

\subsubsection{Effects of Taylor number $Ta$}\label{sec:effects_Ta}

\begin{figure}
	\centering
	\includegraphics[width=0.99\textwidth,trim= 0 0 0 0,clip]{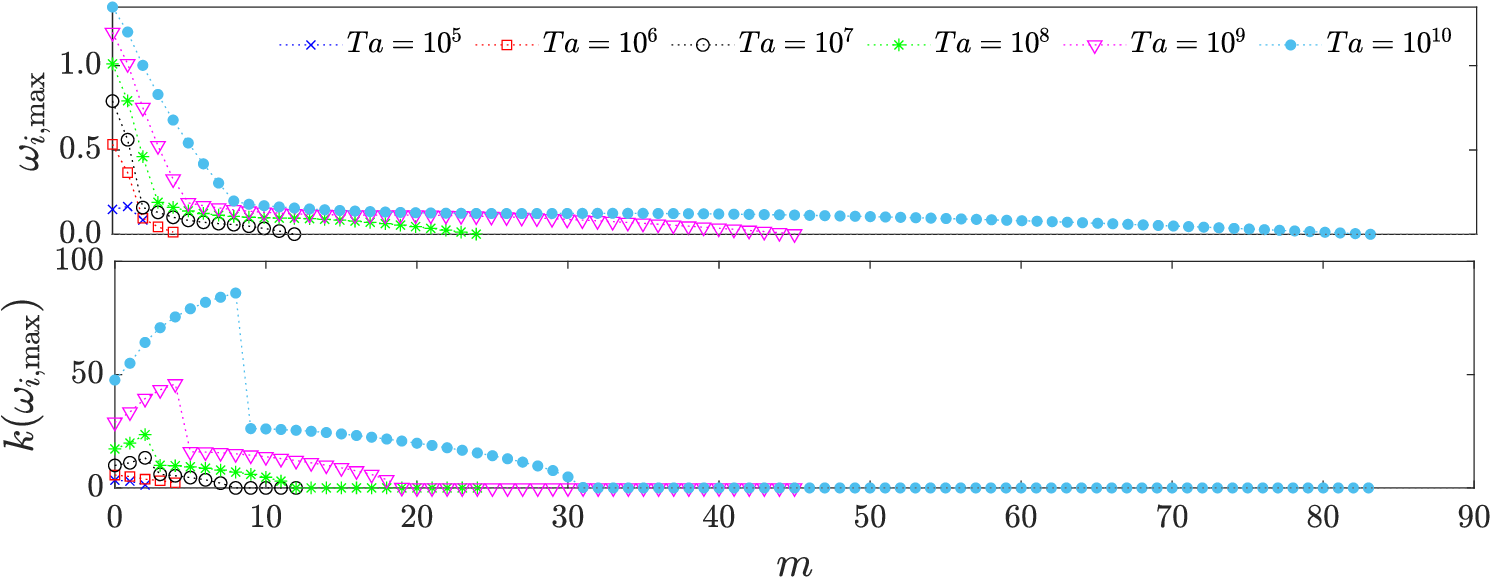}
	\put(-382,137){$(a)$}
	\put(-382,77){$(b)$}			
	\caption{$(a)$ Variations of the largest linear growth rate $\omega_{i,\text{max}}$ with the azimuthal wavenumber $m$ for various Taylor numbers $Ta$ for the quasi-Keplerian flow at $\eta=0.3$, $Pr=0.7$ and $Ri=0.1$. $(b)$ Variations of the corresponding axial wavenumber $k$ at which $\omega_{i,\text{max}}$ is attained.  The discontinuity in the variation of $k(\omega_{i,\text{max}})$ is due to a mode shift from one branch to another branch in the corresponding eigenspectrum. }
	\label{Fig:kmax_Grmax_varyTa}
\end{figure}

In the previous subsection, the Taylor number is primarily fixed at $Ta={10}^6$. Next, we explore the effect of varying $Ta$ while keeping other control parameters fixed in the Keplerian regime at $(q,\eta,Pr,Ri)=(1.5,0.3,0.7,0.1)$.
First, figure \ref{Fig:kmax_Grmax_varyTa}$(a)$ shows that the linear growth rate $\omega_{i,\text{max}}$ of the most unstable disturbance, optimised over all axial wavenumbers $k$, increases monotonically with $Ta$ for any azimuthal wavenumber $m$. This indicates that faster rotation (corresponding to larger $Ta$) enhances the linear instability.
Second, the associated axial wavenumber $k$, plotted in \ref{Fig:kmax_Grmax_varyTa}$(b)$, also increases with $Ta$. This trend features an increasingly thinner mode structure (see again figure 3 of the supplementary material), which aligns more closely with the thin nature of the accretion disks.
Third, the most unstable mode is almost always axisymmetric (except $Ta=10^5$), as shown in panel \ref{Fig:kmax_Grmax_varyTa}$(a)$.

Figure \ref{Fig:kmax_Grmax_varyTa} also shows that as $Ta$ increases, more non-axisymmetric modes become linearly unstable, although they do not dominate. Specifically, for high $m$, the most unstable mode is always axial-independent ($k=0$). Typical non-axisymmetric modes, which exhibit an increasingly localised structure around the outer rotating cylinder, are briefly discussed in section 5 of the supplementary material; see figure 5 therein for the visualisation.
Here, we discuss a potential connection between the observed complex thermal instability diagram and MRI. In this study, the magnetic field is omitted to isolate and focus on thermal effects.
However, in real accretion disks, both magnetic and thermal fields are typically present, even if one or both are relatively weak (see section 4.2.3 of \cite{lesur2022} for a discussion on the coupling of magnetohydrodynamics and thermodynamics in protoplanetary disks). The coexistence of these fields suggests that the thermally-driven linear instability observed in this study may interact---either linearly or nonlinearly---with various manifestations of the MRI, e.g., those in \citep{Hollerbach2005New}. 
Recently, a synergy between thermal effects and magnetic effects on fluid convection has indeed been observed in a Taylor-Couette flow experiment by \cite{Seilmayer2020Convection} and theoretically confirmed in a linear stability analysis by \cite{Mishra2024One}. Note that the gravity considered in their analysis is along the axial direction, unlike our study which models gravity in the radial direction, specifically for accretion disks.


\begin{figure}
	\centering
	\includegraphics[width=0.5\textwidth,trim= 20 0 40 0,clip]{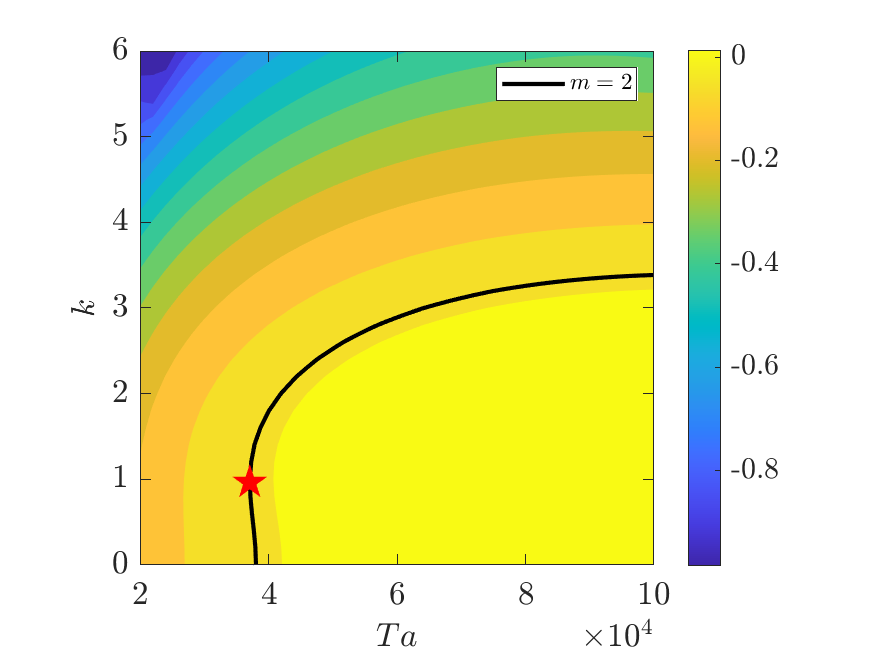}
	\put(-187,150){$(a)$}
	\includegraphics[width=0.425\textwidth,trim= 0 0 0 0,clip]{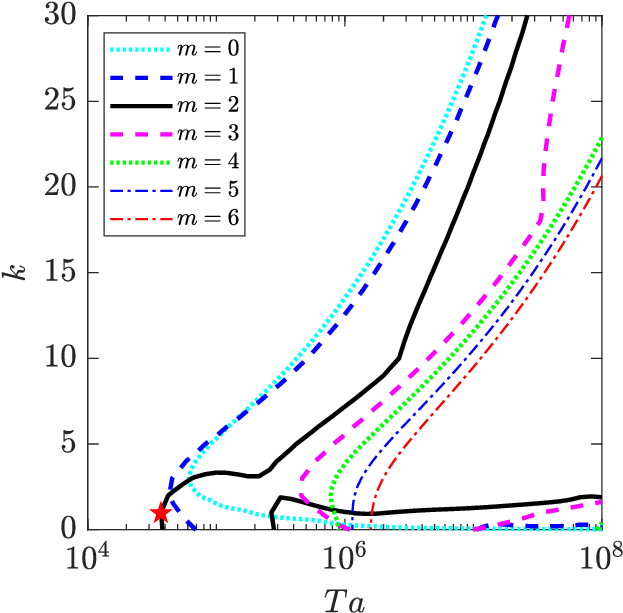}
	\put(-167,150){$(b)$}
	\caption{$(a)$ Contours of the linear growth rate $\omega_{i}$ in the $Ta$-$k$ plane at $m=2$ for the quasi-Keplerian flow at $(\eta,Pr,Ri)=(0.3,0.7,0.1)$. The black curve is the neutral curve for $m=2$. The flow is linearly unstable on the RHS of the curves. $(b)$ Neutral curves at various $m$. Note for $m=(1,3)$ each of the neutral curves consists of two segments, the area confined by which is linearly unstable; the small area at the bottom right corner is linearly stable for $m=(1,3)$ and unstable for $m=(4,5,6)$. Note the log scale for the horizontal axis in panel $(b)$. The red star marks the linear critical condition ($Ta_c\approx3.7\times10^4, k_c\approx0.96, m_c=2$).}
	\label{Fig:NC_m012}
\end{figure}

To determine the critical $Ta_c$, neutral curves along which $\omega_i=0$ are depicted in figure \ref{Fig:NC_m012} for the linearised flow at $(q,\eta,Pr,Ri)=(1.5,0.3,0.7,0.1)$. Panel $(a)$ illustrates contours of the linear growth rate in a $Ta$-$k$ plane, zoomed-in around the linear critical condition ($Ta_c\approx3.7\times10^4, k_c\approx0.96, m_c=2$) marked by the red star and the black neutral curve. For other $m$ values, the Taylor number is always higher than the identified $Ta_c$, as evident in panel $(b)$. Furthermore, it is notable that at $Ta$ significantly above its critical value, there exists a broad spectrum of axial wavelengths of disturbances capable of triggering both axisymmetric and non-axisymmetric linear instabilities. The wider this range and the higher the axial wavenumber, the more relevant the instability becomes to accretion disks.

\subsubsection{Effects of Richardson number $Ri$}

\begin{figure}
	\centering
	\includegraphics[width=0.9\textwidth,trim= 0 35 0 0,clip]{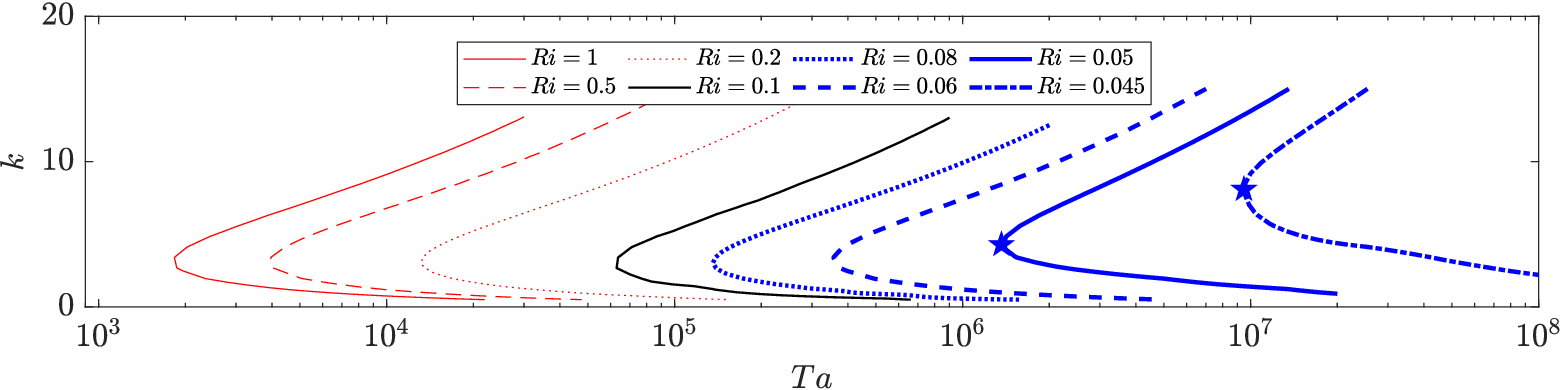}
	\put(-350,60){$(a)$}\put(-300,55){$m=0$}\\
	\includegraphics[width=0.9\textwidth,trim= 0 35 0 0,clip]{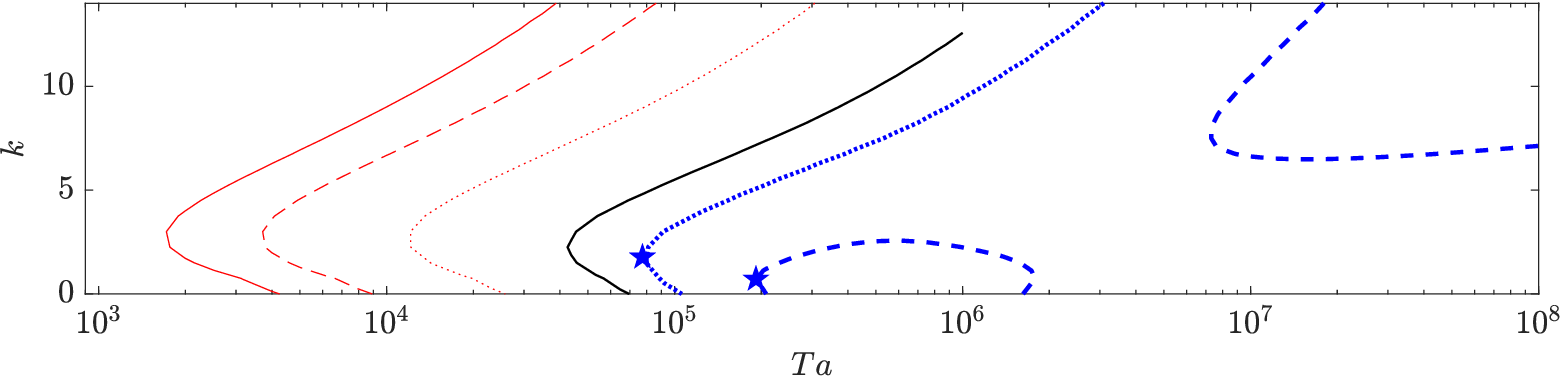}
	\put(-350,56){$(b)$}\put(-300,55){$m=1$}\\
	\includegraphics[width=0.9\textwidth,trim= 0 35 0 0,clip]{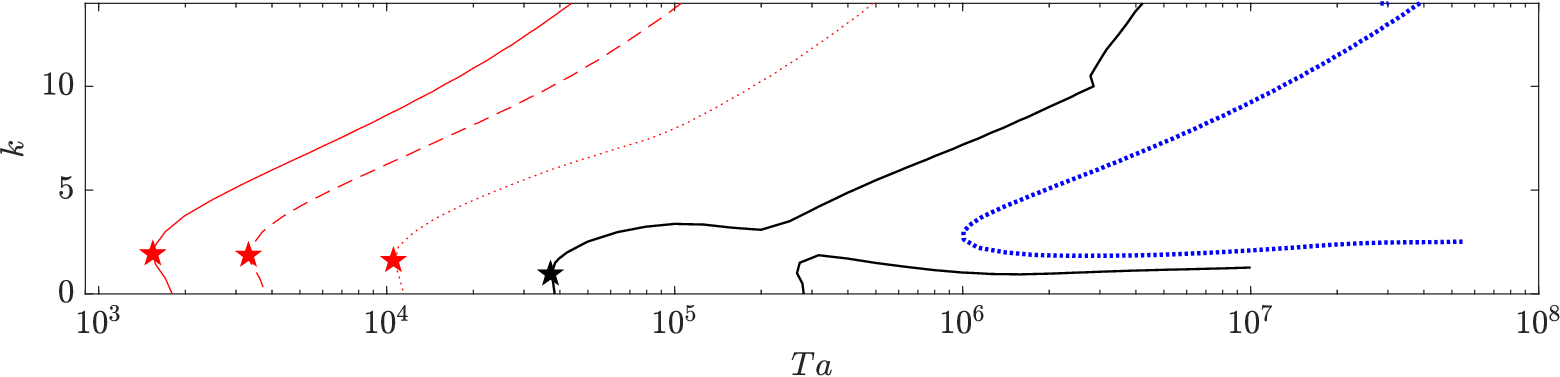}
	\put(-350,56){$(c)$}\put(-300,55){$m=2$}\\
	\includegraphics[width=0.9\textwidth,trim= 0 0 0 0,clip]{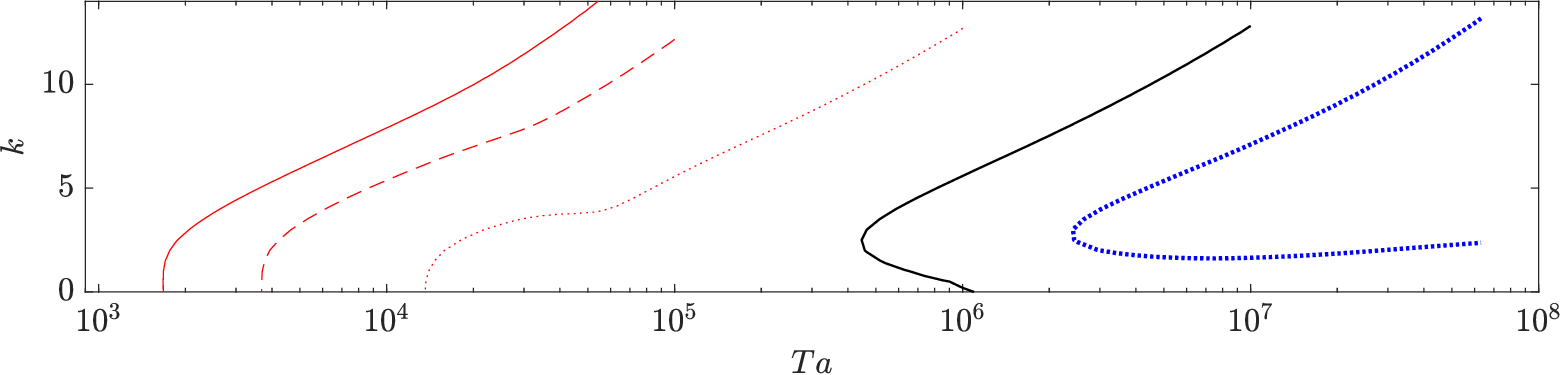}
	\put(-350,75){$(d)$}\put(-300,65){$m=3$}
	\caption{ Neutral curves for various Richardson numbers for the quasi-Keplerian flow at $\eta=0.3$ and $Pr=0.7$: $(a)$ azimuthal wavenumber $m=0$; $(b)$ $m=1$; $(c)$ $m=2$; $(d)$ $m=3$. In all the calculations, only the neutral curve segments close to the left noses are shown and the stars mark the linear critical conditions at each $Ri$ minimised over all $m$ and $k$ values.}
	\label{Fig:NC_varyRi}
\end{figure}

The $Ri$ characterises the thermal effect relative to the shearing in our thermally-driven quasi-Keplerian flow. We find that increasing $Ri$ promotes the linear instability. This can be seen from figure \ref{Fig:NC_varyRi}$(a)$, where neutral curves in the $Ta$-$k$ plane for various $Ri$ are plotted, with parameters $(\eta,Pr)=(0.3,0.7)$ fixed for axisymmetric flows. As our focus is on the linear critical $Ta_c$, only the nose segments of the neutral curves are shown. With increasing $Ri$ from $0.045$ to $1$, the neutral curve shifts from right to left, with the onset $Ta$ decreasing from $10^7$ to $10^3$, four orders of magnitude smaller. 
This suggests that when the destabilising thermal buoyancy effect is strong, the flow is more prone to instability and thus requires a weaker inertial effect, corresponding to a slower rotation, to overcome the stabilising Keplerian shear to trigger the instability.

In addition to the axisymmetric mode, similar destabilising effects of increasing $Ri$ are observed for non-axisymmetric modes; see the neutral curves for $m=(1,2,3)$ in panels $(b,c,d)$, respectively. Certain non-axisymmetric modes become linearly stable, resulting in the absence of neutral curves in the investigated range of $Ta$. Some neutral curves consist of disconnected parts or exhibit kinks, such as $(m,Ri)=(1,0.06)$ (blue dashed lines in panel $b$) and $(m,Ri)=(2,0.1)$ (black curve in panel $c$).
Our result that $Ta_c$ increases with smaller $Ri$ is consistent with the numerical observations in \cite{Held2018Hydrodynamic}, where a simulation study of thermally-stratified quasi-Keplerian flows in a shearing box revealed that weakening the thermal buoyancy results in a linear instability at a larger Rayleigh number. Again, we would like to emphasise that our ``global'' flow configuration is different than their shearing box. 


\begin{figure}
	\centering
	\includegraphics[width=0.99\textwidth,trim= 0 0 0 0,clip]{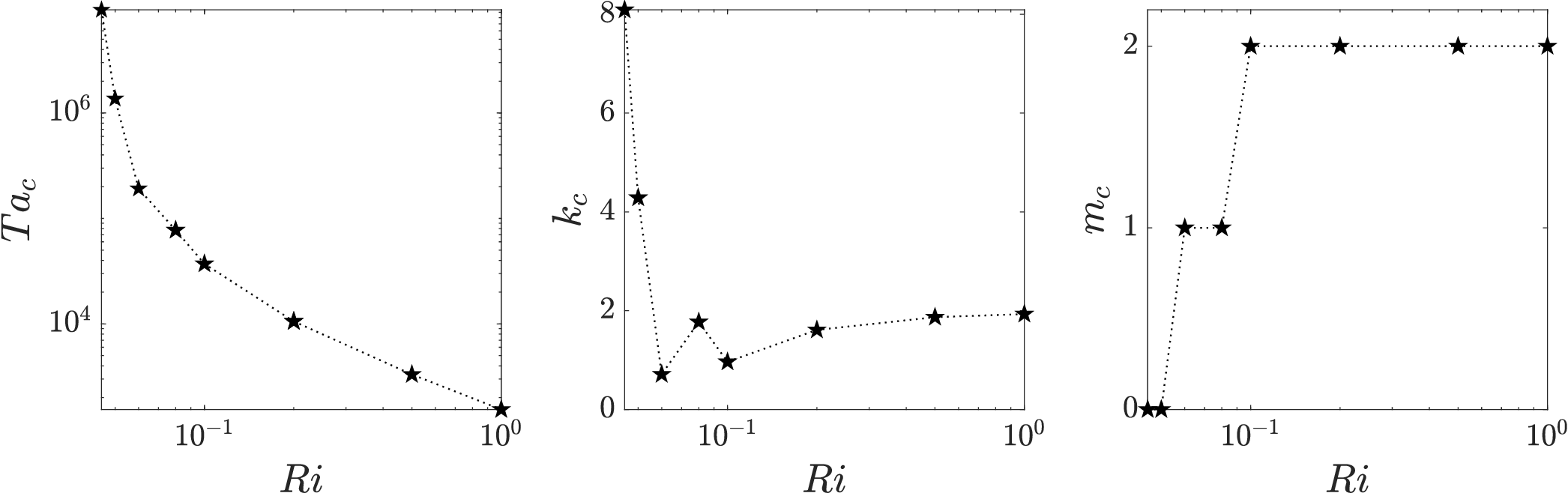}
	\put(-382,111){$(a)$}
	\put(-250,111){$(b)$}
	\put(-121,111){$(c)$}
	\caption{Variations of the linear critical parameters with Richardson number $Ri$ for the quasi-Keplerian flow at $\eta=0.3$ and $Pr=0.7$: $(a)$ linear critical Taylor number $Ta_c$ (see also the stars in figure \ref{Fig:NC_varyRi}); $(b)$ linear critical axial wavenumber $k_c$; $(c)$ linear critical azimuthal wavenumber $m_c$. These data are extracted from the critical points of the neutral curves in figure \ref{Fig:NC_varyRi}$(a$-$c)$; see the eight stars therein.}
	\label{Fig:critical_varyRi}
\end{figure}

\begin{figure}
	\centering
	\includegraphics[width=0.33\textwidth,trim= 0 0 0 0,clip]{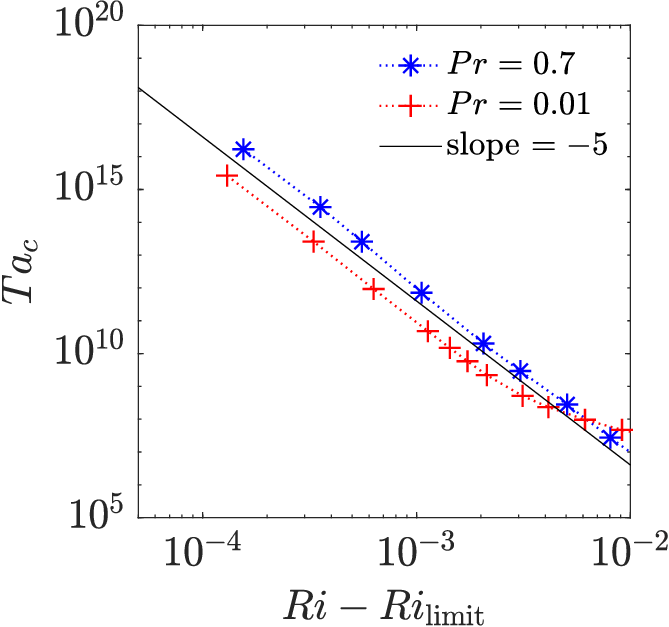}
	\put(-130,111){$(a)$}
	\hspace{0.15in}
	\includegraphics[width=0.33\textwidth,trim= 0 0 0 0,clip]{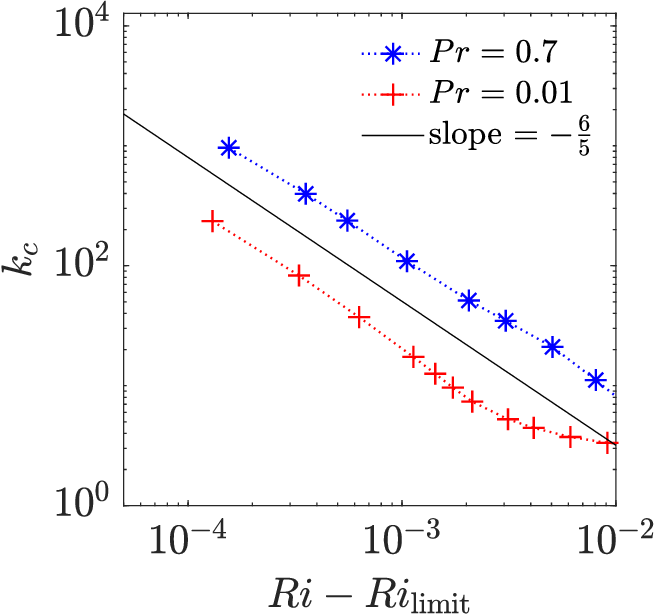}
	\put(-130,111){$(b)$}
	\caption{Scaling laws of $(a)$ linear critical Taylor number $Ta_c  \propto (Ri-Ri_{\text{limit}})^{-5}$ and $(b)$ linear critical axial wavenumber $k_c  \propto (Ri-Ri_{\text{limit}})^{-6/5}$ at the low Richardson number $Ri$ limit for axisymmetric ($m=0$) quasi-Keplerian flow at $\eta=0.3$. The limit values of $Ri_{\text{limit}}$ are computationally estimated to be $Ri_{\text{limit}}\approx 0.034945$ for $Pr=0.7$, and $Ri_{\text{limit}}\approx 0.00087$ for $Pr=0.01$. }
	\label{Fig:critical_smallRi}
\end{figure}

The linear critical $Ta_c$ can be determined by minimising $Ta$ along a neutral curve for all combinations of $(m,k)$ at a given parameter setting. To achieve this, we extract the eight leftmost data points, marked by stars in figure \ref{Fig:NC_varyRi}, and display $(Ta_c,k_c,m_c)$ as functions of $Ri$ in figure \ref{Fig:critical_varyRi}. Here, we observe that $Ta_c$ increases monotonically with reducing $Ri$, with the variation being more pronounced at small $Ri$.
Regarding the linear critical wavenumbers, for relatively high $Ri\geq0.1$, $Ta_c$ are obtained for helical modes with $m_c=2$ and $k_c\approx 2$. In contrast, for relatively small $Ri=(0.05,0.045)$, the linear instability only exists for $m=0$ within the investigated range of Taylor numbers, and $k_c$ dramatically increases upon decreasing $Ri$. These results suggest that when the thermally-driven buoyancy effect is strong, as manifested at relatively large $Ri$, both the axisymmetric and non-axisymmetric disturbances can trigger instabilities, with the non-axisymmetric mode being more dominant at the instability onset. Conversely, for weak buoyancy effect exhibited at relatively small $Ri$, the instability of non-axisymmetric modes seems to be severely suppressed, while the axisymmetric mode instability can still be operative.

The rapid increase of $(Ta_c,k_c)$ at small $Ri$ implies that by further reducing thermal buoyancy through decreasing $Ri$, both the Taylor number and axial wavenumber would increase asymptotically. To explore how these results vary with $Ri$, we look for scaling results in $(Ta_c,k_c)$ with $Ri$ decreasing from $0.05$, as shown in figure \ref{Fig:critical_smallRi}. Our computational results indicate that when $Ri\rightarrow Ri_\text{limit}=0.034945$, the $Ta_c$ approaches infinity under the current numerical resolution at $\eta=0.3, Pr=0.7$. Theoretically, an infinitely large $Ta_c$ means linear stability. As shown in figure \ref{Fig:critical_smallRi}$(a)$, a clear scaling of $Ta_c$ is present with the deviation of $Ri$ from the $Ri_\text{limit}$, as well as the scaling in $k_c$ as a function of $Ri-Ri_\text{limit}$ in panel ($b$).
To further corroborate the scaling result, the case of $Pr=0.01$ is additionally calculated, at which $Ri_\text{limit}\approx0.00087$, showing the same scaling exponents. 
These scaling results may serve as a guide for future high-fidelity numerical investigations of the studied flow. It is noted that similar scalings in the critical parameters have also been observed in other complex fluids as a function of governing parameters, such as viscoelastic fluids, see \cite{Garg2018Viscoelastic}. 

\subsubsection{Effects of Prandtl number $Pr$}

To explore the effects of the Prandtl number $Pr$, the linear critical Taylor number $Ta_c$ is calculated for various $Pr$ covering five orders of magnitude, with other parameters fixed at $(q,\eta,Ri)=(1.5,0.3,0.1)$. Several observations emerge from this analysis.

\begin{figure}
	\centering
	\includegraphics[width=0.33\textwidth,trim= 0 0 0 0,clip]{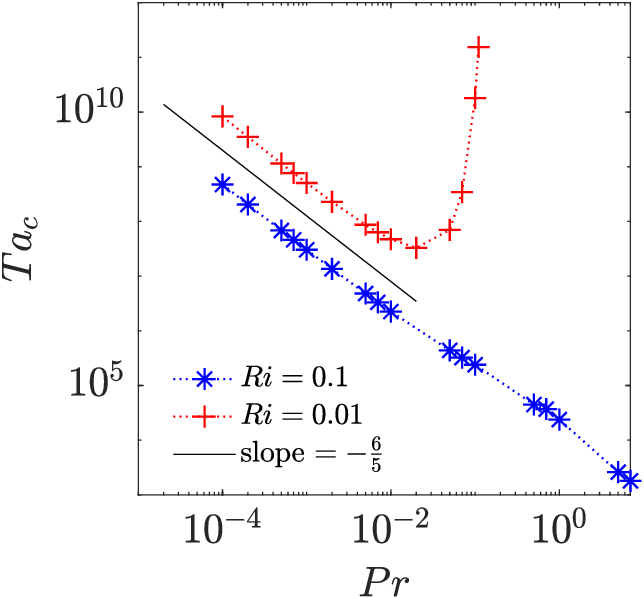}
	\put(-128,114){$(a)$}
	\includegraphics[width=0.32\textwidth,trim= -10 0 0 0,clip]{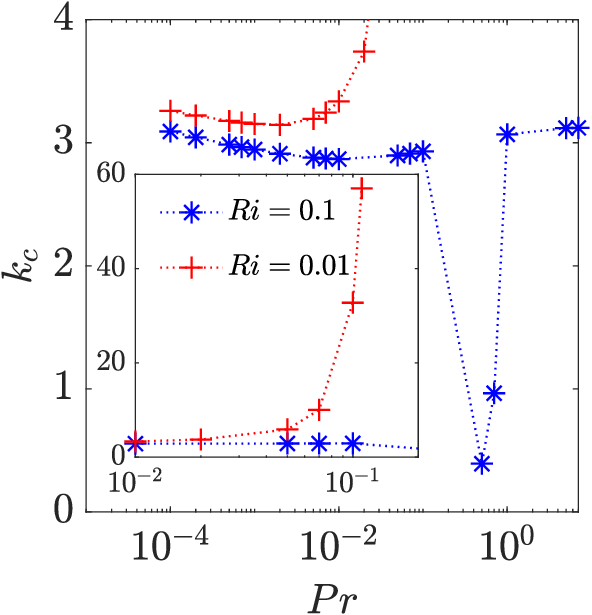}
	\put(-121,114){$(b)$}
	\includegraphics[width=0.32\textwidth,trim= -10 0 0 0,clip]{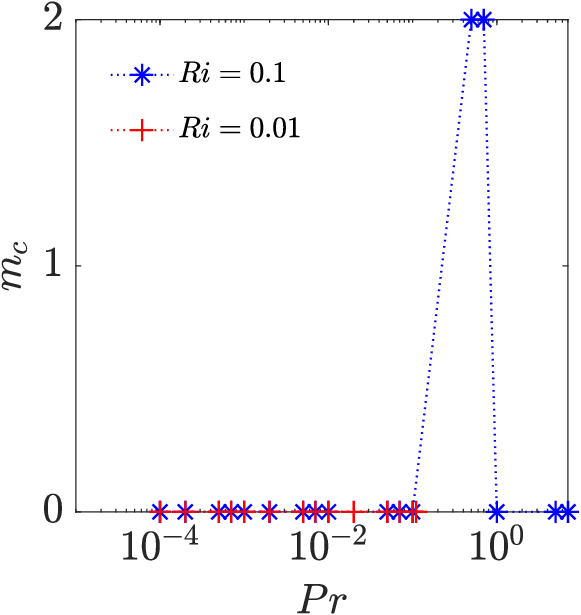}
	\put(-122,114){$(c)$}
	\caption{Variations of the linear critical parameters as functions of Prandtl number $Pr$ for the quasi-Keplerian flow at $\eta=0.3$: $(a)$ linear critical Taylor number $Ta_c$; $(b)$ linear critical axial wavenumber $k_c$; $(c)$ linear critical azimuthal wavenumber $m_c$. Here, $Ta_c \propto Pr^{-6/5}$ when $Pr$ is asymptotically small, but no scaling law in $k_c$ was observed. The inset of panel $(b)$ shows the large variation of $k_c$ when $Ri=0.01$. }
	\label{Fig:NC_varyPr}
\end{figure}

Firstly, the results in figure \ref{Fig:NC_varyPr}$(a)$ reveal that, at $Ri=0.1$, high momentum diffusivity or low thermal diffusivity (meaning greater $Pr$) has destabilising effects on the linear instability in the quasi-Keplerian flow because $Ta_c$ monotonically decreases with increasing $Pr$ from $0.0001$ to $7$.
However, at $Ri=0.01$, the effect of $Pr$ is initially destabilising as it increases from 0.0001 to 0.02, but it becomes strongly stabilising with further increases. The rapid increase of $Ta_c$ suggests a maximal $Pr$ beyond which the viscous dissipation is sufficiently strong to render the flow linearly stable, where the $Ta_c\rightarrow\infty$. In both cases of $Ri=0.1$ and $0.01$, the asymptotic slopes of the function curves approach $-\frac{6}{5}$ in the low-$Pr$ limit, indicating a scaling law of $Ta_c \propto Pr^{-6/5}$. As pointed out by one of the reviewers, caution should be taken with when one attempts to make any link between this trend to the typical scaling of the linear critical Reynolds number $Re_c\propto Pm^{-1}$, where $Pm$ is the magnetic Prandtl number, for the onset of MRI with a pure axial magnetic field in a TC geometry \citep{Rudiger2003Linear,Hollerbach2005New}. Note that $Ta$ is a parameter corresponding to the square of the typical definition of $Re$.

Secondly, figure \ref{Fig:NC_varyPr}$(b)$ shows that the linear critical axial wavenumber $k_c$ mostly remains of $O(1)$ for all $Pr$ values at $Ri=0.1$. This result is consistent with that in figure \ref{Fig:critical_smallRi}$(b)$, where $k_c$ can be extrapolated to $O(1)$ at $Ri=0.1$. Specifically, for intermediate Prandtl numbers $0.5\lessapprox Pr\lessapprox 1$, figure \ref{Fig:NC_varyPr}$(b)$ indicates  $k_c  \sim 1$, meaning that the typical length scale of the these critical modes is comparable to the cylinder gap width. However, for relatively small or large values of $Pr\lessapprox 0.1$ or $Pr\gtrapprox 5$, $k_c$ is approximately equal to $3$.
At $Ri=0.01$, the variation of $k_c$ as a function of $Pr$ presents a radically different behaviour. At the low $Pr$ limit, $k_c$ still remains at about $3$, but when the maximal $Pr$ is approached, $k_c$ increases dramatically, see the red crosses in the inset of panel ($b$). Accompanying the scaling law of $Ta_c$ at the low $Pr$ limit, the variation of $k_c$ is small, increasing by less than $10\%$ when $Pr$ is reduced by two orders of magnitude. No observable scaling law of $k_c$ is evident within the investigated range of Prandtl number.

Note that in both $Ri$ cases, $k_c\sim 3$ at the small-$Pr$ limit. A similar value of $k_c\approx3.13$ has been reported by \cite{Jenny2007Primary} in a geophysical study based on the TC flow modelling. Likewise, \cite{Meyer2021Stability} documented an asymptotic wavenumber of about $k_c\approx3.1$ at large $Pr$ or $Ra$ for a TC flow with centrifugal buoyancy effects. The close value of $k_c$ in these different TC configurations suggests the generic features shared by the various flow instabilities in TC flows subjected to radial buoyancy due to stratification.
It should be emphasised that the above (relatively) long-wavelength feature pertains to the linear critical conditions, whereas the short-wavelength modes, which is more astrophysically relevant, will dominate in the quasi-Keplerian flow for $Ta\gg Ta_c$, as illustrated in figure \ref{Fig:kmax_Grmax_varyTa}.

Lastly, it is found that axisymmetric modes dominate in the linear instability at high and low $Pr$. As illustrated in figure \ref{Fig:NC_varyPr}$(c)$, for the $Pr$ values studied at $Ri=0.1$, $Ta_c$ is consistently attained at $m=0$, except for the cases $Pr=(0.5,0.7)$, for which $Ta_c$ is obtained for helical modes with $m=2$. Thus, non-axisymmetric modes are only significant at the instability onset in a narrow range of $Pr$.
At $Ri=0.01$, axisymmetric mode dominates the instability onset at all the $Pr$ investigated.

\subsubsection{Effects of radius ratio $\eta$}\label{sec:radius_ratio_effect}

In the preceding analysis, the radius ratio of the co-rotating cylinders is held fixed at $\eta=0.3$. However, the relevant radius ratio in the TC flow to model the actual accretion disks may be much smaller than $0.3$ due to their inherently flat and thin structures around small central objects \citep{Abramowicz2014Accretion}. In this section we investigate how varying $\eta$ affects the linear instability in the thermal-TC flow.

\begin{figure}
	\centering
	\includegraphics[width=0.99\textwidth,trim= 0 0 0 0,clip]{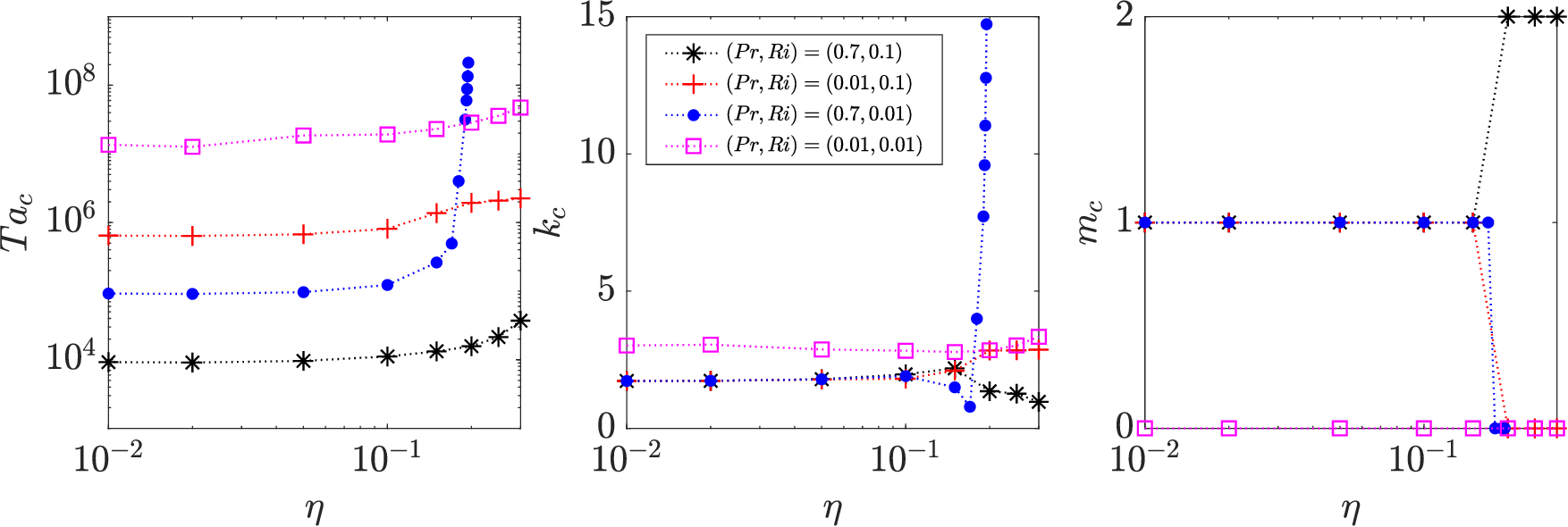}
	\put(-381,119){$(a)$}
	\put(-253,119){$(b)$}
	\put(-123,119){$(c)$}
	\caption{Variations of the linear critical parameters as functions of radius ratio $\eta$ for the quasi-Keplerian flow at different $(Pr,Ri)$ settings: $(a)$ linear critical Taylor number $Ta_c$; $(b)$ linear critical axial wavenumber $k_c$; $(c)$ linear critical azimuthal wavenumber $m_c$.}
	\label{Fig:critical_varyeta}
\end{figure}

Figure \ref{Fig:critical_varyeta}$(a)$ reveals that reducing the radius ratio destabilises the quasi-Keplerian flow for the four selected combinations of $(Pr,Ri)$. Specifically, for the case at $(Pr,Ri)=(0.7,0.01)$, no linear instability can be found when $\eta\gtrapprox 0.2$. This demonstrates the significant influence of the curvature effect on the quasi-Keplerian flow stability/instability. In contrast, \cite{Meyer2021Stability} observed that reducing radius ratio stabilised quasi-Keplerian flows subjected to centrifugal buoyancy in the TC geometry with a higher temperature at the outer cylinder---a configuration distinct from ours.

Panel \ref{Fig:critical_varyeta}$(b)$ depicts the variation of the corresponding linear critical $k_c$, showing that the onset of instability is consistently triggered by disturbances with finite axial wavelength. This axial dependence suggests that the linear instability sets in with axial variation, at least when $Ta$ is slightly beyond $Ta_c$.  Regarding the linear critical $m_c$, panel \ref{Fig:critical_varyeta}$(c)$ illustrates that at small $\eta\lessapprox 0.1$ the instability onset is  consistently dominated by the first helical mode at $m=1$, except for the case at $(Pr,Ri)=(0.01,0.01)$, which is dominated by axisymmetric mode.
At relatively large $\eta$, the case $(Pr,Ri)=(0.7,0.1)$ favours $m_c=2$, whereas all the other cases pick the axisymmetric mode.

\subsubsection{A comparison to the local stability analysis of the flow}\label{app:comparison_local}

As suggested by one of the reviewers, we next compare the results of our stability analysis to those of a local analysis \citep{Klahr2014Convective,Lyra2014Convective}. In a local analysis, flow homogeneity has been assumed in all the directions. To distinguish, we will call our analysis as global, which is inhomogeneous in the radial direction. 

In the following, we select control parameters in the local analysis as close as possible to those used in sections \ref{sec:effects_Ta} to \ref{sec:radius_ratio_effect}. The obtained local results also reveal a linear instability in the thermally-stratified quasi-Keplerian flow. The observed effects of various governing parameters on this instability are also qualitatively consistent to those in the global analysis. Differences lie in two aspects. First, the local analysis based on the short wave approximation only provides limited information on the instability growth rate due to spatially periodic disturbances of short wavelength; it gives no information on the unstable mode pattern. In contrast, the global analysis uncovers rich mode patterns of the instability, exhibiting various length scales in all three directions; see figure \ref{Fig:axisymmetric_mode}, figure \ref{Fig:velo_eta005_PrRi1em3} and more in the supplementary material. Second, the instability onset Taylor number and axial wavenumber from the two analyses show notable differences; and so do the corresponding scaling laws. All the comparisons, either qualitative or quantitative, are summarised in table \ref{Tab:compare_local_global}. Below, we describe them in detail.

\begin{table}
	\begin{center}
		\begin{tabular}{p{0.4cm}  p{3.4cm} |p{2.9cm} p{2.9cm} p{2.9cm}}
			No.  & Quantities compared & Global results & Local results& Parameters \\ \hline
			1  & The highest linear growth rate $\omega_i$ and the corresponding axial wavenumber & $\omega_i\approx1.35$, $k\approx47.62$ & $\omega_i\approx1.05$, $k_z\approx67$ at $r_0=0.5$  & $(\eta,Pr,Ri,Ta)$= $(0.3,0.7,0.1,10^{10})$  \\
			2  & Linear critical Taylor number & $Ta_c \approx 6\times 10^4$  & $Ta_c \approx 5\times 10^5$&$(\eta,Pr,Ri)$= $(0.3,0.7,0.1)$ \\
			& & & & \\
			3 & Effect of increasing $Ri$ & Destabilising &  Destabilising & $(\eta,Pr)$=$(0.3,0.7)$ \\
			4  & The scaling law of the linear critical parameters & $Ta_c\propto (Ri-Ri_{\text{limit}})^{-5}$, $Ri_{\text{limit}}\approx 0.034945$; $k_c\propto (Ri-Ri_{\text{limit}})^{-6/5}$  & $Ta_c\propto (Ri-Ri_{\text{limit}})^{-1}$, $Ri_{\text{limit}}\approx 0.0453873$; $k_{z,c}\approx4.44$ & $(\eta,Pr)$=$(0.3,0.7)$\\
			& & & & \\
			5 & Effect of increasing $Pr$ & Destabilising &  Destabilising & $(\eta,Ri)$=$(0.3,0.1)$ \\
			6 & The scaling law of the linear critical parameters & $Ta_c\propto Pr^{-6/5}$; \, \, \, \, $k_{c}\approx3$  & $Ta_c\propto Pr^{-1}$; \, \, \, \,  $k_{z,c}\approx4.44$ & $(\eta,Ri)$=$(0.3,0.1)$ \\
			& & & & \\
			7  & Effect of reducing $\eta$ & Generally destabilising& Generally destabilising & $(Pr,Ri)$=$(0.01,0.01)$\\
			8  & The scaling law of the linear critical parameters & No scaling law of $Ta_c$; $k_{c}\approx3$  & No scaling law of $Ta_c$; $k_{z,c}\approx4.44$ & $(Pr,Ri)$=$(0.01,0.01)$\\ 
		\end{tabular}
		\caption{Comparisons of the results from the global analysis (presented in section \ref{sec:results}) and the local stability analysis (in this appendix) of the same quasi-Keplerian flow in the TC geometry. For all the comparisons, $m=0$. For No. 2 to 8 in the comparisons, the local results are obtained at a fixed radial location of $r_0=(r_i+r_o)/2$.} 
		\label{Tab:compare_local_global}
	\end{center}
\end{table}

\begin{figure}
	\centering
	\includegraphics[width=0.99\textwidth,trim= 0 0 0 0,clip]{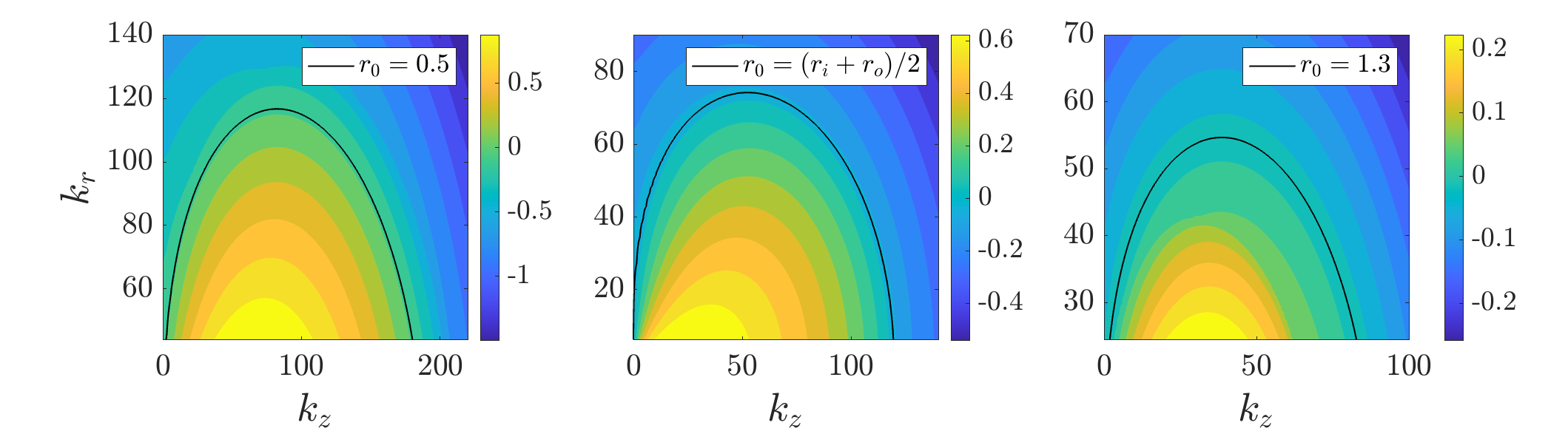}
	\put(-368,94){$(a)$}\\
	\includegraphics[width=0.43\textwidth,trim= 0 0 -20 0,clip]{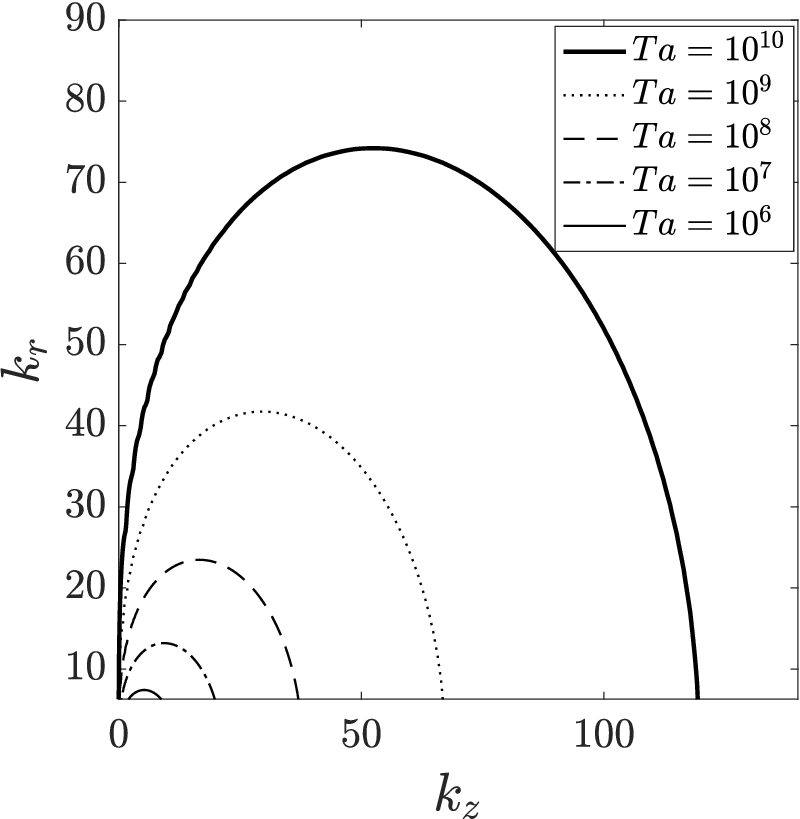}
	\put(-167,151){$(b)$}
	\includegraphics[width=0.445\textwidth,trim= 0 0 0 0,clip]{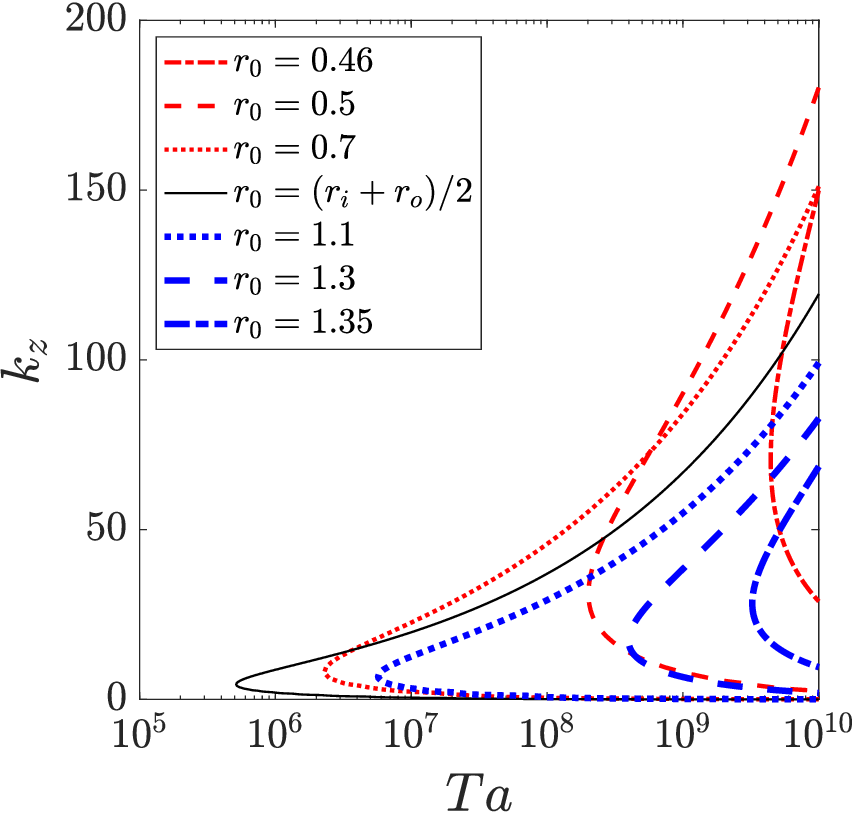}
	\put(-171,151){$(c)$}\\
	\includegraphics[width=0.93\textwidth,trim= 0 0 0 0,clip]{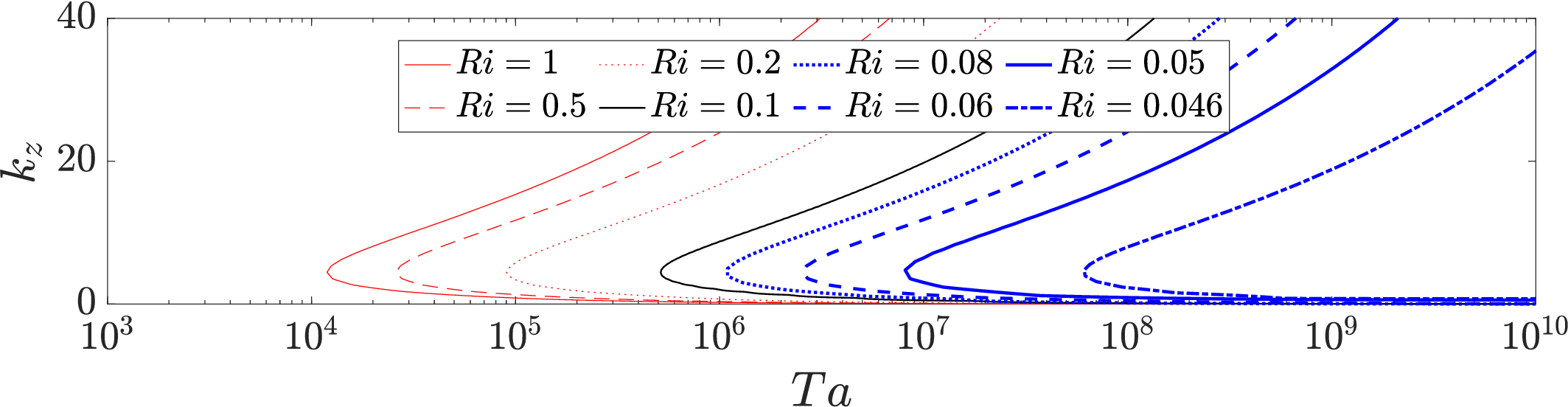}
	\put(-358,84){$(d)$}	
	\caption{Linear instability observed in the local analysis of the quasi-Keplerian flow at $(\eta,Pr,Ri,m)=(0.3,0.7,0.1,0)$. $(a)$ Contours of the linear growth rate in the wavenumber plane for the flow at three different radial locations $r_0$ and $Ta=10^{10}$. The solid black curve represents the neutral curve. $(b)$ Variation of the neutral curve in the wavenumber plane with $Ta$ for the local flow at the center of the cylinder gap $r_0=(r_i+r_o)/2$. $(c)$ Variation of the neutral curve in the $Ta$-$k_z$ plane with varying radial locations $r_0$. $(d)$ Variation of the neutral curve with Richardson number $Ri$ for the local flow at the center of the cylinder gap $r_0=(r_i+r_o)/2$.}
	\label{Fig:local_analysis}
\end{figure}

In all the calculations for the local analysis, solving the $5\times 5$ eigenvalue problem \ref{eq:perturbation_spec_shortwave} results in five eigenvalues, from which the leading eigenvalue is always selected for extracting the linear growth rate.
Figure \ref{Fig:local_analysis}$(a)$ displays contours of the linear growth rate $\omega_i$ in the wavenumber plane at three different radial locations for the quasi-Keplerian flow at $(\eta,Pr,Ri,Ta,m)=(0.3,0.7,0.1,10^{10},0)$. The area below the neutral curve indicates linear instability obtained from the local analysis. The corresponding instability in the global analysis is reported in figure \ref{Fig:kmax_Grmax_varyTa}, where the largest growth rate is $\omega_i\approx1.35$ at $(k,m)=(47.62,0)$. In figure \ref{Fig:local_analysis}$(a)$, the largest growth rate of $\omega_i\approx1.05$ is attained at $(k_r,k_z)\approx(44,67)$ for $r_0=0.5$. Quantitatively, both the growth rate and the axial wavelength are of the same order of magnitude. The radial location $r_0=0.5$ also aligns with the localised mode pattern depicted in figure 3$(b)$ of the supplementary material.  Another important observation is that for a given $k_z$, the growth rate increases as $k_r$ decreases, indicating that the local instability favours relatively large radial structures. Comparatively, the global analysis is better suited for resolving large-scale unstable modes.

Figure \ref{Fig:local_analysis}$(b)$ shows how the neutral curve varies with $Ta$ in the wavenumber plane for the local flow at the center of the cylinder gap $r_0=(r_i+r_o)/2$. The expansion of the neutral curve with increasing $Ta$ indicates a destabilising effect, consistent with the global analysis reported in section \ref{sec:effects_Ta}. However, this variation is specific to the radial location $r_0=(r_i+r_o)/2$. To determine the critical $Ta$ which depends on $r_0$, we plot neutral curves in the $Ta$-$k_z$ plane for various values of $r_0$ in figure  \ref{Fig:local_analysis}$(c)$. In this plot, the radial wavenumber $k_r$ is set to its lowest allowed value at each radial location $r_0$, i.e., $k_r=\text{max}[\pi/(r_0-r_i),\pi/(r_o-r_0)]$ (note that for a given radial wavenumber $k_r$, the disturbance wavelength is $L_r=2\pi/k_r$ and the whole wave must be confined within the radial domain, requiring that $r_0-L_r/2\ge r_i$ and $r_0+L_r/2\le r_o$). This choice of $k_r$ ensures the highest linear growth rate maximised over $k_r$. Figure \ref{Fig:local_analysis}$(c)$ shows that $r_0=(r_i+r_o)/2$ corresponds to the smallest critical $Ta$ above which the local instability occurs, with an approximate value of $Ta_c\approx 5\times10^{5}$. This value is of a similar order of magnitude to that observed in figure \ref{Fig:NC_m012}$(b)$ from the global analysis.

\begin{figure}
	\centering
	\includegraphics[width=0.99\textwidth,trim= 0 0 0 0,clip]{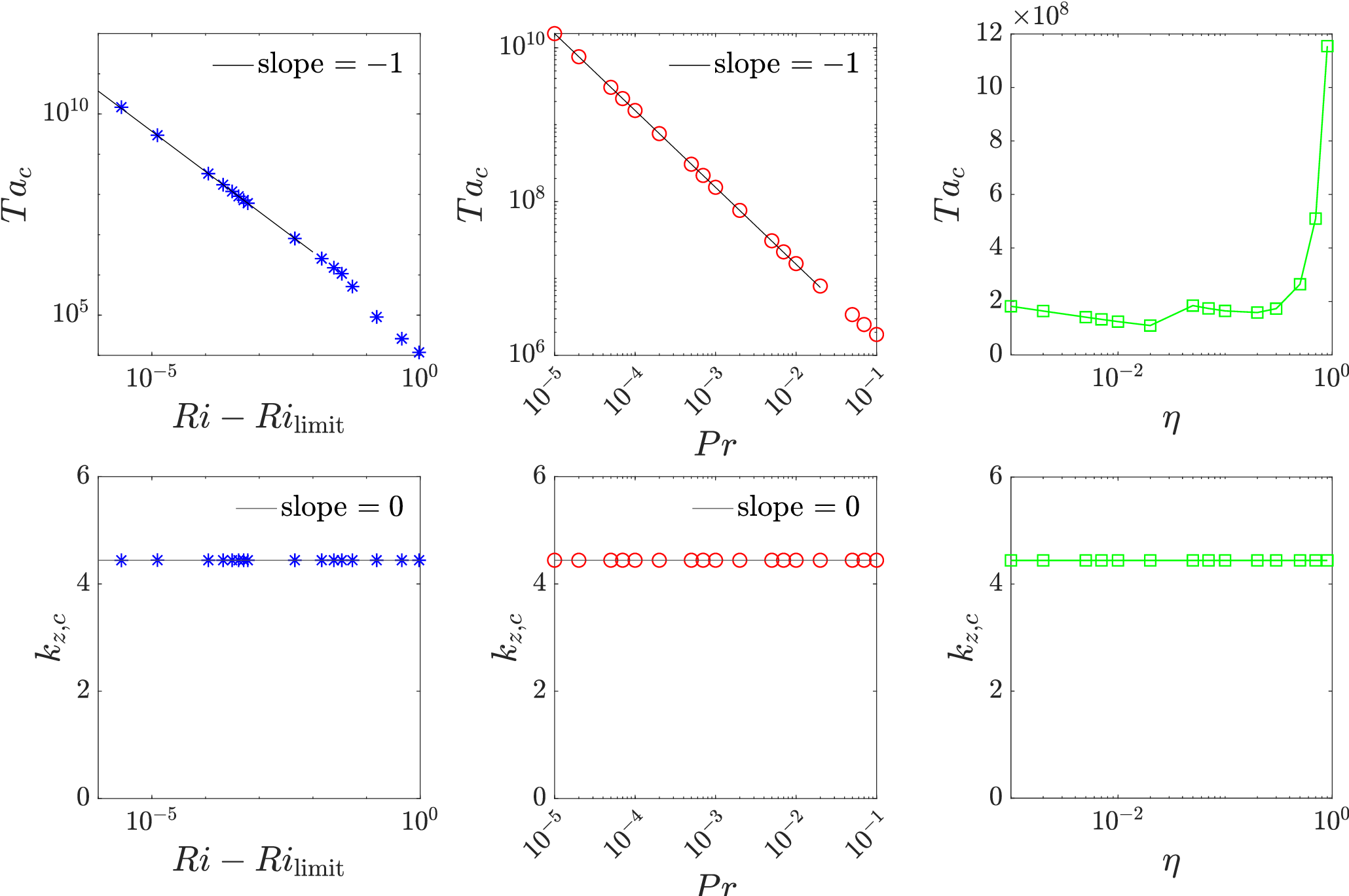}
	\put(-382,235){$(a)$}
	\put(-255,235){$(b)$}
	\put(-120,235){$(c)$}
	\put(-382,110){$(d)$}
	\put(-255,110){$(e)$}
	\put(-120,110){$(f)$}	
	\caption{Variations of the linear critical Taylor number $Ta_c$ and the linear critical axial wavenumber $k_{z,c}$ with the control parameters $(Ri,Pr,\eta)$ for the axisymmetric ($m=0$) quasi-Keplerian local flow at the cylinder gap center $r_0=(r_i+r_o)/2$. $(a,d)$ At the low Richardson number $Ri$ limit and $(\eta,Pr)=(0.3,0.7)$. The limit values of $Ri_{\text{limit}}$ are computationally estimated to be $Ri_{\text{limit}}\approx 0.0453873$. $(b,e)$ At the small Prandtl number $Pr$ limit and $(\eta,Ri)=(0.3,0.1)$. $(c,f)$ At varying viscosity ratio $\eta$ and $(Pr,Ri)=(0.01,0.01)$. In panels $(d,e,f)$, $k_c\approx 4.44$. }
	\label{Fig:local_scaling}
\end{figure}
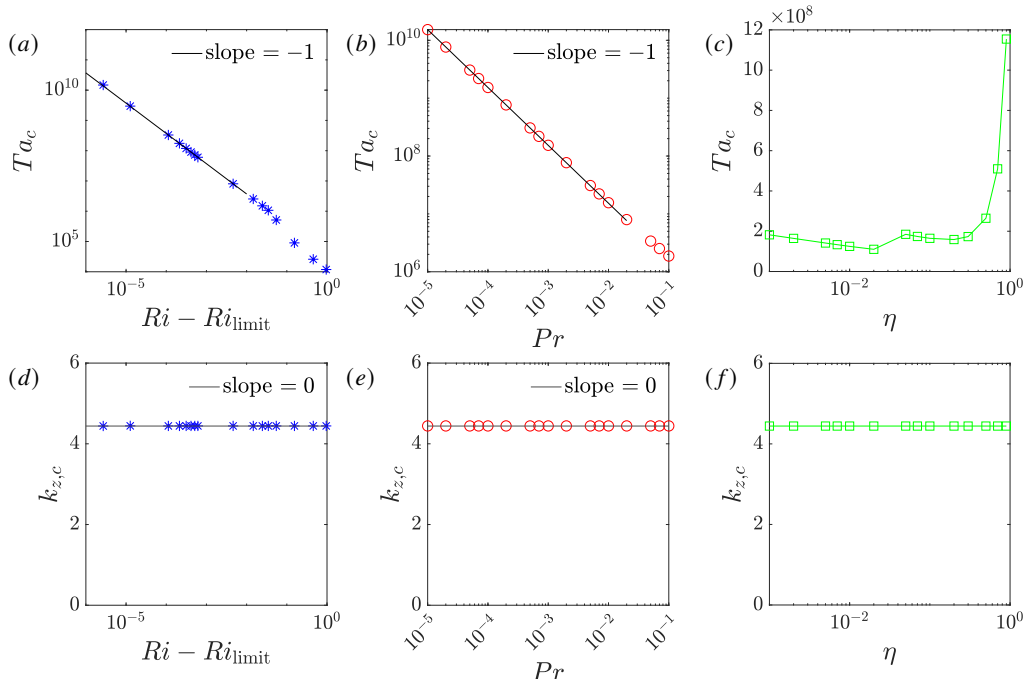

In light of the above effect of $r_0$ on the local instability, we fix $r_0=(r_i+r_o)/2$ to study the influence of various governing parameters on the instability onset. Figure \ref{Fig:local_analysis}$(d)$ illustrates the destabilising effect of increasing the Richardson number $Ri$ on the neutral curve, consistent with the global analysis results presented in figure \ref{Fig:NC_varyRi}$(a)$. However, a notable difference is that the linear critical axial wavenumber $k_{z,c}$ remains constant across varying $Ri$; also see figures \ref{Fig:local_scaling}$(b,d,f)$, where it maintains a value of approximately $4.44$ despite changes in $(\eta,Pr,Ri)$. The reason for this constant value remains unclear. Figure \ref{Fig:local_scaling}$(a)$ shows $Ta_c$ as a function of $Ri$. It can be seen that the scaling law of $Ta_c$  has significantly changed, with an exponent of $-1$ compared to $-5$ in the global analysis. Increasing the Prandtl number $Pr$ is also destabilising, as evidenced by figure \ref{Fig:local_scaling}$(c)$, where the scaling law is $Ta_c\propto Pr^{-1}$. This is in close agreement with the scaling $Ta_c\propto Pr^{-6/5}$ from the global analysis shown in figure \ref{Fig:NC_varyPr}$(a)$. Additionally, figure \ref{Fig:local_scaling}$(e)$ demonstrates that reducing the radius ratio $\eta$ generally has a destabilising effect, consistent with the trend observed in figure \ref{Fig:critical_varyeta}$(a)$ from the global analysis.

In a nutshell, the results from the local stability analysis of the quasi-Keplerian flow are qualitatively consistent with those from the global analysis, indicating that similar underlying physics is at play. However, quantitatively, the scaling laws of the linear critical conditions differ between the local and global analyses, highlighting the distinctiveness of these two approaches. Comparatively, the global analysis provides more solid results than the local analysis. The local analysis involves numerous assumptions and the large-scale mode structures resolved in the global analysis can never be reproduced in the local analysis, underscoring the limitations of the local approach.

\subsection{Linear instability at extreme parameters}\label{sec:extreme_case}

The preceding subsection demonstrates that generally, decreasing $Pr$ or $Ri$ has stabilising effects, while decreasing $\eta$ is destabilising. To discuss the values of these parameters relevant to accretion disks in the literature, we mention that \cite{Held2018Hydrodynamic} used $Pr=2.5$ and $Ri=0.05$ in their numerical simulations of thermal quasi-Keplerian flows in a 3-D local shearing box. \cite{Latter2016Convective} noted that $Pr$ could be as low as $10^{-7}$ in the inner radii and the author adopted a small $Ri$ value of $0.01$. For a TC geometry, appropriate values of $\eta$ are not directly extractable from the literature due to the absence of an ``outer rotating cylinder'' in real accretion disks. For modeling purposes, one may consider small values since disks are typically thin and flat with relatively small central objects \citep{Abramowicz2014Accretion}. This subsection illustrates that the above reported thermally-driven linear instability persists with the control parameters becoming as small as $(\eta,Pr,Ri)=(0.05,10^{-3},10^{-3})$, which are more astrophysically relevant. The linear instability at much more severe parameters is further demonstrated in section 5 of the supplementary material.
In addition, the influence of the gravitational acceleration profile is examined at the parameter setting of $(\eta,Pr,Ri)=(0.05,10^{-3},10^{-3})$ and discussed in Appendix \ref{app:gravity_profile}, showing that a stronger gravity field enhances the linear instability and vice versa.

\subsubsection{The critical condition for the flow at $(\eta,Pr,Ri)=(0.05,10^{-3},10^{-3})$}

\begin{figure}
	\centering
	\includegraphics[width=0.99\textwidth,trim= 30 0 30 0,clip]{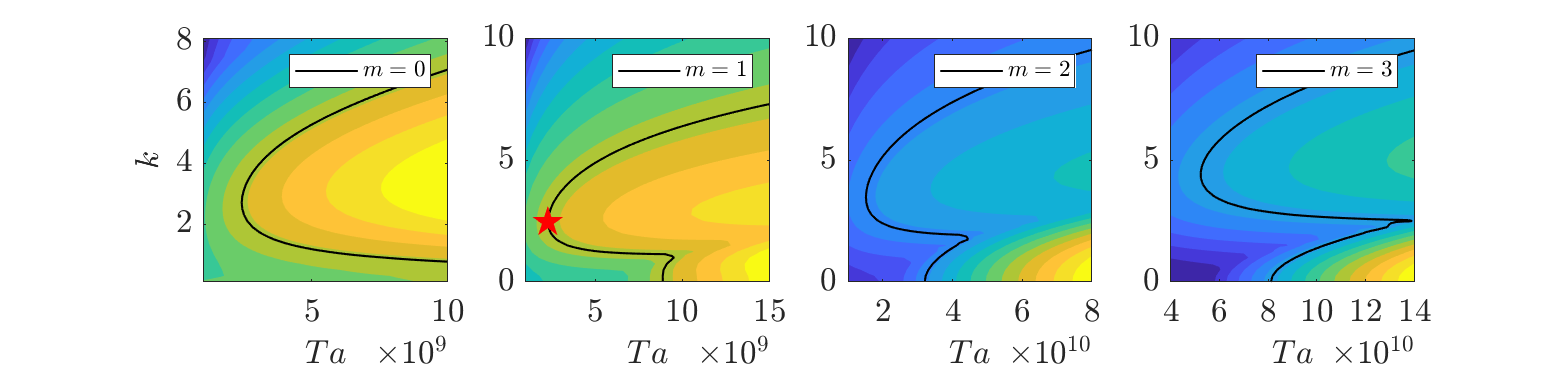}
	\put(-363,82){$(a)$}
	\put(-275,82){$(b)$}
	\put(-189,82){$(c)$}
	\put(-105,82){$(d)$}
	\caption{Contours of the linear growth rate $\omega_{i}$ in the $Ta$-$k$ plane for the quasi-Keplerian flow at $(\eta,Pr,Ri)=(0.05,10^{-3},10^{-3})$. The black curves are the neutral curves ($\omega_i=0$) and the red star marks the linear critical condition at about $(Ta_c,k_c,m_c)=(2.28 \times 10^{9},2.46,1)$.  }
	\label{Fig:NC_eta005_m0123}
\end{figure}

\begin{figure}
	\centering
	\includegraphics[width=0.99\textwidth,trim= 0 0 0 0,clip]{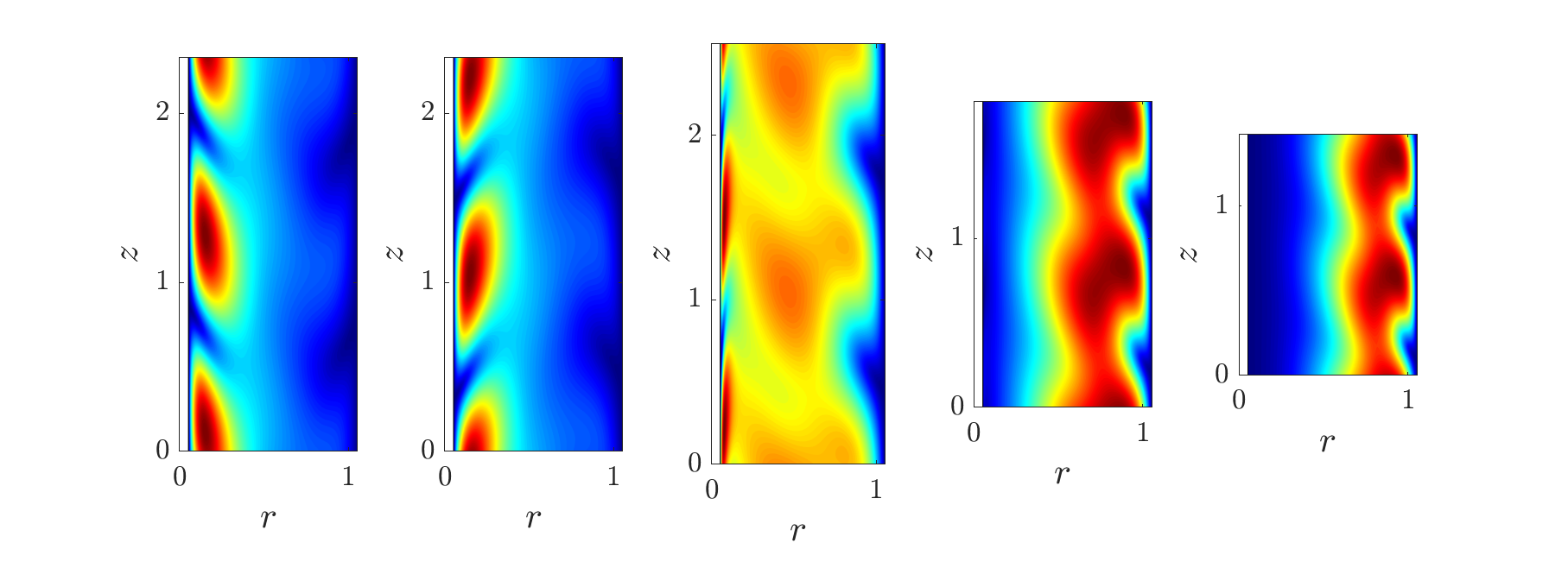}
	\put(-360,115){$(a)$}
	\put(-325,130){$m=0$}
	\put(-260,130){$m=0$}
	\put(-197,130){$m=1$}
	\put(-132,128){$m=2$}
	\put(-69,128){$m=3$}\\
	\includegraphics[width=0.99\textwidth,trim= 0 0 0 0,clip]{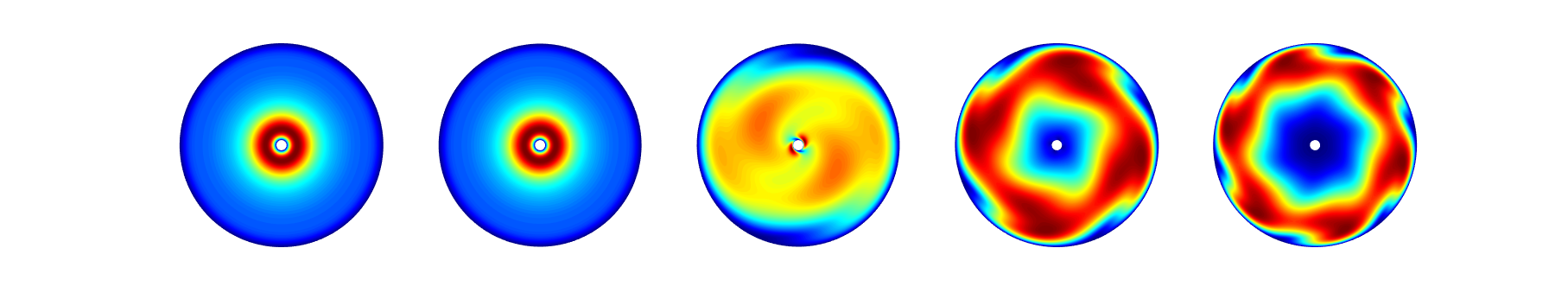}
	\put(-360,35){$(b)$}
	\caption{Contours of the perturbation velocity magnitude at the four critical points identified in figure \ref{Fig:NC_eta005_m0123}. Panel $(a)$ shows the cross sections in the $r$-$z$  plane where only one wavelength is plotted at the azimuthal angle $\theta=0$; panel $(b)$ shows the cross sections in the $r$-$\theta$  plane at the axial position $z=0$. From left to right, the first and second panels in each row correspond to a pair of complex conjugate modes at $m=0$, the third to fifth panels are for $m=1,2,3$, respectively. }
	\label{Fig:velo_eta005_PrRi1em3}
\end{figure}
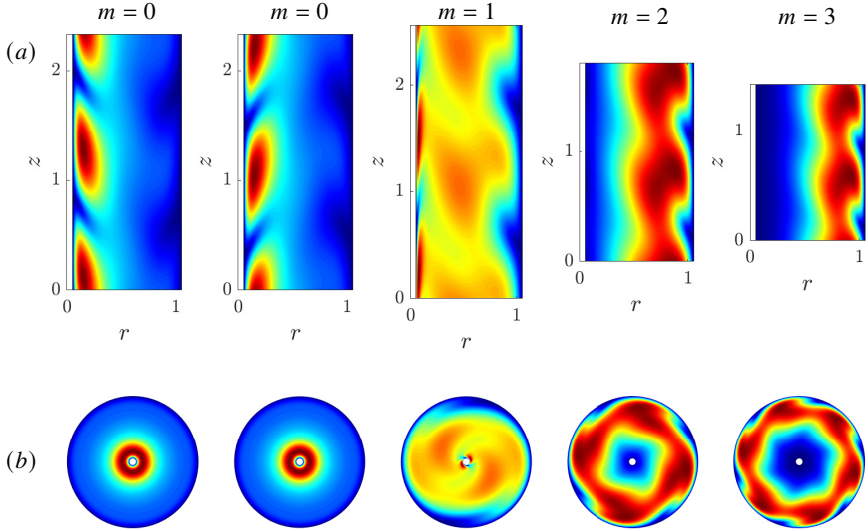

The neutral curves in the case $(\eta,Pr,Ri)=(0.05,10^{-3},10^{-3})$ for different values of $m$ are shown in figure \ref{Fig:NC_eta005_m0123}, from which one can see that the linear instability sets in at approximately $(Ta_c,k_c,m_c)=(2.28 \times 10^{9},2.46,1)$, see the red star in panel ($b$). The eigenvalue of this neutral mode at the critical condition is $\omega\approx-0.15607945375-0.00000000002i$. The frequency corresponds to a negative phase speed of $c_r=\omega_r/m\approx -0.156$, indicating that this mode rotates in the opposite direction of the base flow when viewed in the rotating frame of reference. After transforming back to the fixed frame of reference (see section 3 of the supplementary material for a description of the transformation method), its phase speed becomes $c_r=[\omega_r + m/(2Ro)]/m=\omega_r/m + 1/(2Ro)\approx 0.0587$, much smaller than the lowest angular frequency $\omega_o =1/(2Ro)\approx 0.214$ of the base flow, which is attained at the outer rotating cylinder. The pattern of this mode at $m=1$ is visualised in the third column of figure \ref{Fig:velo_eta005_PrRi1em3}($a,b$), showing that the disturbance fills the entire cylinder gap, with the largest variations occurring in the thin vicinity near the inner rotating cylinder. In the $r$-$z$ plane, the vortices attached to the inner cylinder are strongly tilted, indicating that the mode travels in the negative $z$ direction when viewed in the rotating frame of reference. This is consistent with its phase speed in the $z$ direction $c_r=\omega_r/k_c\approx -0.0634$.

In addition to the helical mode at $m=1$, the axisymmetric mode ($m=0$) warrants examination due to the proximity of its neutral curve's lowest $Ta$ value ($2.41 \times 10^{9}$) to $Ta_c\approx 2.28 \times 10^{9}$. This closeness suggests potential interaction between these modes in nonlinear simulations of flow transition at the instability onset. The neutral mode has a pair of complex conjugate eigenvalues $\omega\approx \pm 0.307448512+0.000000003i$. The structures of this pair of modes are visualised in the first two columns of figure \ref{Fig:velo_eta005_PrRi1em3}($a,b$). In the $r$-$z$ plane, localised vortices are seen tilting up and down near the inner rotating cylinder, indicating that these two modes propagate at the same speed but in opposite directions. In the $r$-$\theta$ plane, the structure is identical when visualised at the axial location $z=0$, as shown in the first two circular figures in panel \ref{Fig:velo_eta005_PrRi1em3}($b$). In addition, the last two columns of figure \ref{Fig:velo_eta005_PrRi1em3}($a,b$) depict the flow structures of the non-axisymmetric modes $m=(2,3)$ at the identified critical points. As $m$ increases, the disturbance becomes more localised and moves closer to the outer rotating cylinder. Similar behaviour can be observed in figure 5 of the supplementary material. It should be noted that the parameters used here and those for figure 5 therein are significantly different, suggesting that the shifted localisation of the mode with increasing $m$ may be a generic feature across the parameter space.

\subsubsection{Beyond the critical condition: linear instability at $Ta={10}^{16}$} \label{sec423}

With the instability onset identified at $(Ta_c,k_c,m_c)=(2.28 \times 10^{9},2.46,1)$, the following question arises: is this critical condition relevant to accretion disks? A key factor in this consideration may be the axial wavelength of the mode. In this case, the axial wavelength is on the order of the cylinder gap, making it incompatible with accretion disks because the characteristic axial length scale in real disks is usually much smaller than the radial length scale, such as the disk radius \citep{Latter2016Convective}. To obtain more relevant results, we have searched for linear instabilities with very small axial wavelengths. Indeed, figure \ref{Fig:kmax_Grmax_varyTa} demonstrates that linear instability persists at Taylor numbers significantly higher than the onset value ($Ta\gg Ta_c$). At such high $Ta$, the linear instability spans a wide range of high axial wavenumbers, corresponding to small axial length scales, which are more similar to the accretion disk conditions.

\begin{figure}
	\centering
	\includegraphics[width=0.90\textwidth,trim= 0 0 0 0,clip]{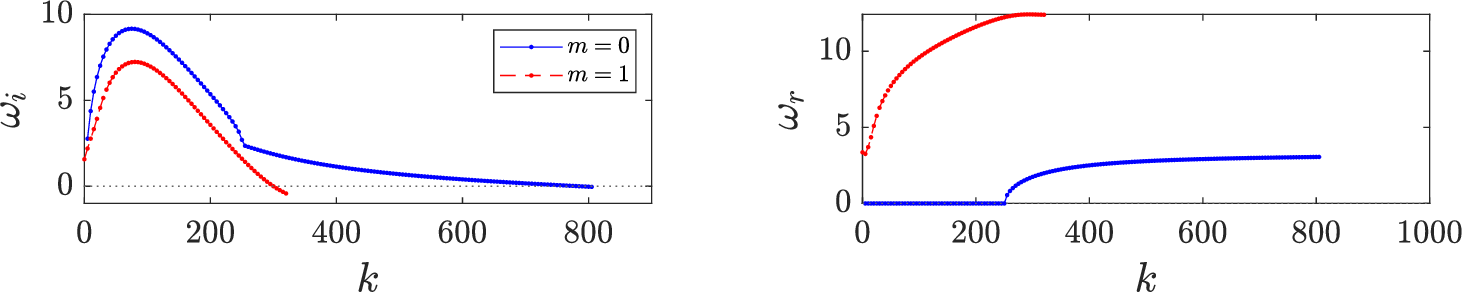}
	\put(-349,60){$(a)$}\put(-165,60){$(b)$}
	\caption{ ($a$) Linear growth rate $\omega_i$ as function of $k$ for the quasi-Keplerian flow at $Ta=10^{16}, \eta=0.05$, $Pr=10^{-3}$ and $Ri=10^{-3}$. $(b)$ Frequency $\omega_r$ as function of $k$ for the same flow.}
	\label{Fig:eta005_Ta1e16_m01}
\end{figure}

For the case at $(\eta,Pr,Ri)=(0.05,10^{-3},10^{-3})$, we investigate the unstable mode at $Ta=10^{16}$, seven orders of magnitude larger than the critical value of about $10^9$. The highest $Ta$ previously investigated in Taylor-Couette flows is about $10^{13}$ \citep{Grossmann2016High}. Although $Ta=10^{16}$ may be challenging to achieve in numerical simulations and physical experiments, it is feasible in our linear stability analysis provided sufficient spatial resolution in the spectral method. For the following calculations, we use $401$ Chebyshev-Lobatto nodes. Figure \ref{Fig:eta005_Ta1e16_m01}$(a)$ shows the dispersion relation $\omega_i(k)$ for the case abovementioned. At this $Ta$, the most unstable mode becomes the axisymmetric mode ($m=0$), rather than the helical mode at $m=1$ which dominates at the linear critical condition, as discussed in figure \ref{Fig:NC_eta005_m0123}$(b)$. Notably, the flow is linearly unstable over a wide spectrum of axial wavenumbers $k$. For $m=0$, the instability range is $0< k<800$; for $m=1$, the range narrows but remains broad at $0\le k<300$. Higher azimuthal wavenumbers $m>2$ also exhibit linear instability, but with much smaller linear growth rates, so they are not shown. The non-smoothness of the blue curve for $m=0$ in figure \ref{Fig:eta005_Ta1e16_m01}$(a)$ at around $k\approx 250$ is due to the transition of the most unstable mode from stationary to oscillatory, as seen in the frequency variations $\omega_r(k)$ in panel $(b)$.
Lastly, it should be mentioned that in addition to the above most unstable mode at $m=0$ and the leading unstable mode at $m=1$, there are a magnitude of axisymmetric and non-axisymmetric unstable modes in the corresponding eigenspectra for this case, signifying the high complexity of the instability diagram. As these modes are less unstable, we will not characterise them further.

\section{Conclusions}\label{sec:conclusion}

Motivated by research efforts searching for hydrodynamic instabilities without magnetic effects that may be operative in protoplanetary disks, we conducted a three-dimensional ``global'' analysis of thermally-stratified quasi-Keplerian flows between a hot inner rotating cylinder and a cold outer rotating cylinder.
Unlike previous work by \cite{Meyer2021Stability}, our model incorporates stellar gravity in the radial direction and considers the temperature distribution within accretion disks, which decreases with increasing distance from the central star.
The flow instability in the TC geometry should also be distinguished from that found based on short wave approximations \citep{Klahr2014Convective,Lyra2014Convective} , as the latter bears only local significance.

We identified a robust flow instability in the quasi-Keplerian flow regime driven purely by radial thermal stratification along with radial gravity, without the need for a magnetic field as required by MRI. This instability manifests in both infinitesimal axisymmetric and non-axisymmetric modes across two or three dimensions, exhibiting typical unstable eigenmode structures resembling those in canonical Rayleigh-B{\'e}nard convection but distorted by the laminar circular Couette flow, especially in non-axisymmetric cases. Our calculation of the linear critical conditions ($Ta_c,k_c,m_c$) revealed that decreasing the Prandtl number $Pr$ or Richardson number $Ri$ generally stabilises the quasi-Keplerian flow. Conversely, reducing the radius ratio $\eta$ is destabilising. At small values of  $Pr=0.001$ and $Ri=0.001$,
the linear instability persists at sufficiently large $Ta\gtrsim 10^9$ with small axial wavelength in the weakly thermally-stratified quasi-Keplerian flow. This suggests that even very weak thermal buoyancy is sufficient to trigger the linear instabilities. Likewise, \cite{Balbus1991Powerful} showed that even a weak magnetic field suffices to cause MRI.
Additionally, we revealed two scaling laws. The first one shows that $Ta_c\propto Pr^{-6/5}$ at the small-$Pr$ limit, implying again the relevance to accretion disks where $Pr$ can be as low as $10^{-7}$ \citep{Latter2016Convective}. The second one reveals that $Ta_c\propto (Ri-Ri_{\text{limit}})^{-5}$, indicating the stabilising effect of decreasing $Ri$. These scaling laws may help to guide future high-fidelity numerical investigations or experiments of the studied flow. 

This thermally-driven hydrodynamic linear instability and subsequent nonlinear dynamics may provide a mechanism for explaining radially outward angular momentum transport in unmagnetised accretion disks or dead zones where MRI is inoperative. We note that \cite{Lyra2014Convective} has successfully produced self-sustained 3-D vortex flows from the saturation of a linear overstability in the modelling of protoplanetary disks; however, their linear instability is localised since they assumed homogenisation in three directions and the evolution may need to be further investigated in a global sense. In addition, \cite{Held2018Hydrodynamic} have demonstrated a series of accretion-disk dynamics from the linear instability (of axisymmetric mode only) to turbulence based on a shearing box approximation of the quasi-Keplerian flow also with thermal stratification but in the axial direction. Unlike these previous localised linear instability assumptions, our 3-D global analysis complements existing discussions, furthering our understanding of accretion-disk dynamics by showing that even very weak radial thermal stratification can cause sufficiently strong gravitational buoyancy to drive the quasi-Keplerian flow linearly unstable.
Meanwhile, we point out that the TC flow is a limited model for the accretion disk. One of the drawbacks is the configuration of two no-slip no-penetration cylinder walls which do not exist in real accretion disks. This work focuses on the momentum transport mechanism in accretion disks. Nevertheless, we hope that the present work on the radial thermal stratification could inspire more studies along this vein of research and contribute to a deeper understanding of the complex accretion disk dynamics. Future research avenues may include investigating flow bifurcation beyond the linear instability, conducting direct numerical simulations of nonlinear development, exploring turbulent flows in thermally-driven TC flow, and assessing the effects of fluid compressibility on linear instability.


\backsection[Acknowledgements]{ We thank all the reviewers for their constructive suggestions.}

\backsection[Funding]{DW was supported by a PhD scholarship (No. 201906220200) from the China Scholarship Council and an NUS research scholarship. MZ acknowledges the financial support of NUS (Suzhou) Research Institute and National Natural Science Foundation of China (Grant No. 12202300) and the MOE Tier 1 grant A-8001172-00-00. We also acknowledge the financial support from the Max-Planck Society, the Deutsche Forschungsgemeinschaft through grants 521319293 and 540422505, and the Daimler and Benz Foundation.}

\backsection[Declaration of interests]{The authors report no conflict of interest.}




\appendix

\section{Illustration of key differences between the current work and two previous studies on the same topic}\label{app:two_studies}

In the introduction section, we briefly mentioned the links of our study to some relevant prior works. Below, we describe in detail the differences between our study and \cite{Meyer2021Stability} and \cite{Klahr2014Convective} to demonstrate the position and novelty of our research.

\cite{Meyer2021Stability} similarly conducted the linear stability analysis of the quasi-Keplerian flow in a Taylor-Couette geometry. However, their flow configuration, with a cold inner cylinder and a hot outer cylinder, does not align with the temperature distribution of accretion disks, where the temperature decreases in the radial direction \citep{Fromang2019Angular}. Additionally, their focus was on the effects of centrifugal buoyancy, neglecting gravitational acceleration, which may not be directly applicable for modelling accretion disks under stellar gravity. Moreover, the parameters most relevant to disk dynamics comprise low $Pr$ \citep{Latter2016Convective} and high $Ta$ \citep{Ji2006Hydrodynamic,Grossmann2016High}, which is at variance with the high Prandtl number and low Taylor number considered in \cite{Meyer2021Stability}. The present study, however, takes these perspectives into account. More detailed comparisons are shown in table \ref{Tab:compare_Meyer2021PRF}. Despite the differences mentioned, we acknowledge that the general destabilising mechanism in our study and that in \cite{Meyer2021Stability} are similar: density stratification due to a thermal gradient can be unstable under acceleration from external forces when the acceleration is in the opposite direction to the density stratification. Nevertheless, we believe that our study adds value to the explanation of angular momentum transport in accretion disks with more detailed results.

Regarding the comparison with \cite{Klahr2014Convective}, there are three main differences. First, they modelled the thermal relaxation process, where the heating and cooling timescales play a crucial role, in their discovery of the convective overstability. This is more realistic concerning the thermodynamics in accretion disks than our thermal equilibrium assumption, which was adopted for simplicity. Second, their analysis was local, performed in a small volume around a given spatial location $(r,\theta)$ (note that they ignored the axial velocity $u_z$ equation). This local assumption excludes the possibility of any linear instability triggered by large-scale structures. In contrast, our model is ``global'', with flow boundaries in the radial direction and the entire $2\pi$ radian in the azimuthal direction. Third, they made various assumptions and simplifications in deriving the dispersion relation due to the local model selection. In contrast, we numerically solve a generalised eigenvalue problem for the complete dispersion relation. Table \ref{Tab:compare_Klahr2014} shows more detailed comparisons.

\begin{table}
	\begin{center}
		\begin{tabular}{c c c}
			Items compared & \cite{Meyer2021Stability} & The present study \\ \hline
			Thermally-stratified TC flow & Yes & Yes \\
			Stratification direction & Radially & Radially \\			
			Boussinesq approximation & Yes & Yes \\
			Incompressible & Yes & Yes \\			
			Inner cylinder temperature & Cold&Hot\\
			Outer cylinder temperature & Hot&Cold\\
			Gravitational force & Ignored& Incorporated\\
			Centrifugal force & Incorporated& Ignored\\
			Destabilising mechanism & Centrifugal buoyancy & Gravitational buoyancy \\
			Prandtl number regime focused & High $Pr$: $[1,10^3]$ & Low $Pr$: $[10^{-4}, 10]$ \\
			Rayleigh number regime focused & $Ra$: $[10^3,10^5]$ & Equivalent to $Ra$: $[10^{3}, 10^{10}]$ \\
			Richardson number regime focused & N.A. & Low $Ri$: $[10^{-3}, 10^{-1}]$ \\						
			Taylor number regime focused & Low $Ta$: $Ta\lessapprox 10^3$ & High $Ta$: $[10^{6}, 10^{16}]$
		\end{tabular}
		\caption{ Differences and similarities between the present study and \cite{Meyer2021Stability}. }
		\label{Tab:compare_Meyer2021PRF}
	\end{center}
\end{table}

\begin{table}
	\begin{center}
		\begin{tabular}{c c c}
			Items compared & \cite{Klahr2014Convective} & The present study \\ \hline
			Thermally-stratified TC flow & Yes & Yes \\
			Stratification direction & Radially & Radially \\
			Thermal relaxation process & Modelled& Assuming thermal equilibrium \\
			Model type & Local shearing box & Global 3-D model \\
			Incompressible & Yes & Yes \\			
			Inviscid & Yes & No\\
			Gravitational force & Ignored& Incorporated\\
			Centrifugal force & Ignored& Ignored\\
			Continuity equation & Neglected& Solved\\
			Axial velocity & Ignored &  Solved\\
			WKB approximation & Yes &  No\\
			Destabilising mechanism & Thermal relaxation driven & Gravitational buoyancy
		\end{tabular}
		\caption{ Differences and similarities between this study and \cite{Klahr2014Convective}. }
		\label{Tab:compare_Klahr2014}
	\end{center}
\end{table}

In summary, both of the two abovementioned works are relevant to accretion disks, though the modelling levels differ. Subsequent works on the convective overstability by \cite{Lyra2014Convective}, \cite{Latter2016Convective} and \cite{Held2018Hydrodynamic} followed the same local approximation in a shearing box, so comparisons will not be repeated here.

\section{Influence of the gravitational acceleration profile on the linear instability}\label{app:gravity_profile}

In the present study, the gravitational acceleration profile $g(r)=r_o/r$ is suited for the Taylor-Couette flow geometry described in cylindrical coordinates. However, for realistic accretion disks described in spherical coordinates, the profile due to the central object should follow $g(r)=r_o^2/r^2$. To the best of our knowledge, neither of these profiles has been realisable in laboratory experiments to date. This implies that numerical simulations are likely the most viable approach to further investigate the complex nonlinear dynamics of the observed linear instability.
Fortunately, a constant profile $g(r)=1$ (non-dimensional), has been realised in experiments on thermo-electric flow convection in a cylindrical annulus \citep{Antoine2023Thermo}. This ``electric gravity'' can be generated due to dielectrophoretic forces when an electric field is applied radially. To minimise the impact of Earth's gravity on the experimental setup, these experiments can be conducted during sounding rocket flights, which provide a microgravity environment \citep{Antoine2023Thermo}. Although the ``electric gravity'' technique is expensive and challenging, it offers a potential avenue for investigating the present thermally-driven linear instability in quasi-Keplerian flows. Future work in this direction could benefit from linear stability analysis using the constant profile $g(r)=1$. As suggested by one of the reviewers, in this appendix we compare the effects of the above three different gravitational acceleration profiles on the linear instability.

These profiles can be expressed collectively as
\begin{equation}
g(r;p) = \frac{r_o^p}{r^p} \, \text{with} \, p=0,1,2.
\end{equation}
At the outer rotating cylinder $r_o$, the non-dimensional gravitational acceleration is always equal to one, as $g_o^*$ is chosen as the reference value for non-dimensionalisation. For the flow field at $r<r_o$, $g(r)\ge 1$. The maximum value of $g(r)$ is attained at the inner rotating cylinder $r_i$, with $g(r_i;p=2)>g(r_i;p=1)>g(r_i;p=0)=1$.

\begin{figure}
	\centering
	\includegraphics[width=0.45\textwidth,trim= 0 0 0 0,clip]{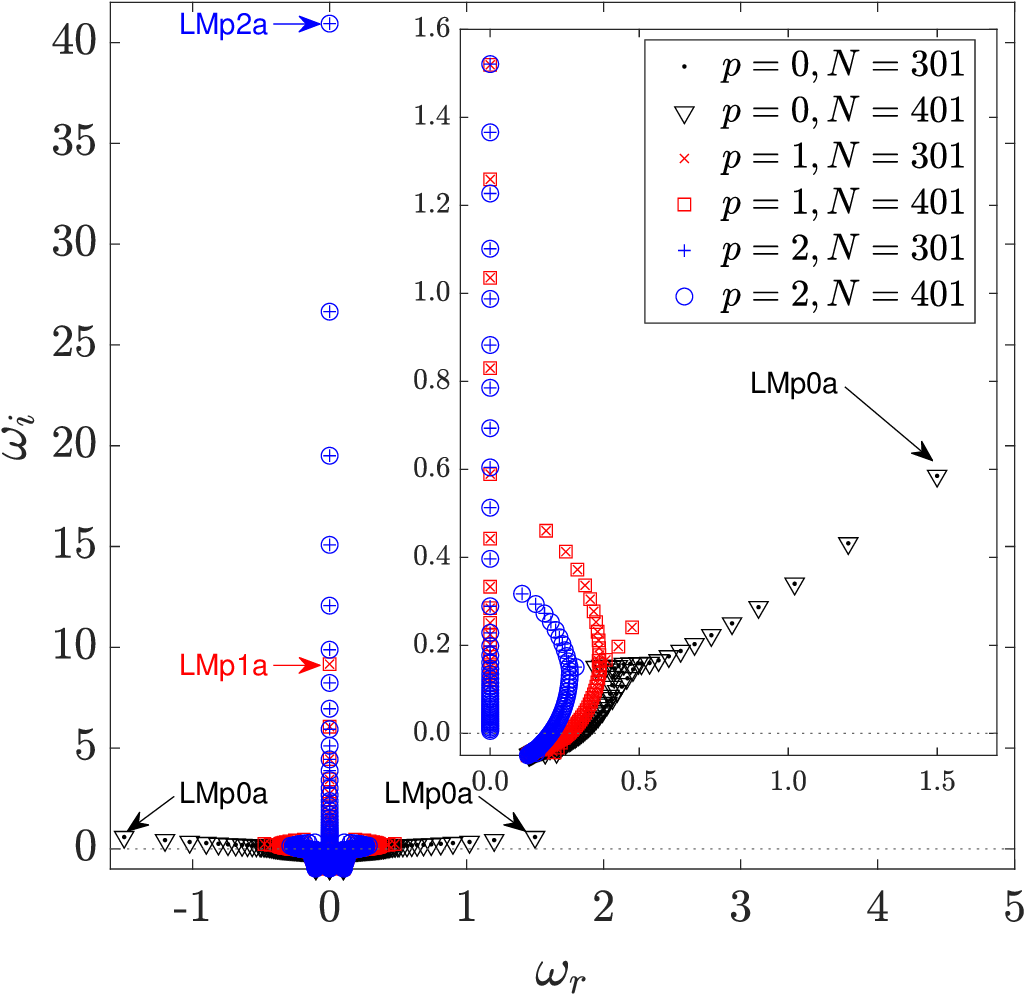}
	\put(-180,160){$(a)$}\hspace{0.2cm}
	\includegraphics[width=0.45\textwidth,trim= 0 0 0 0,clip]{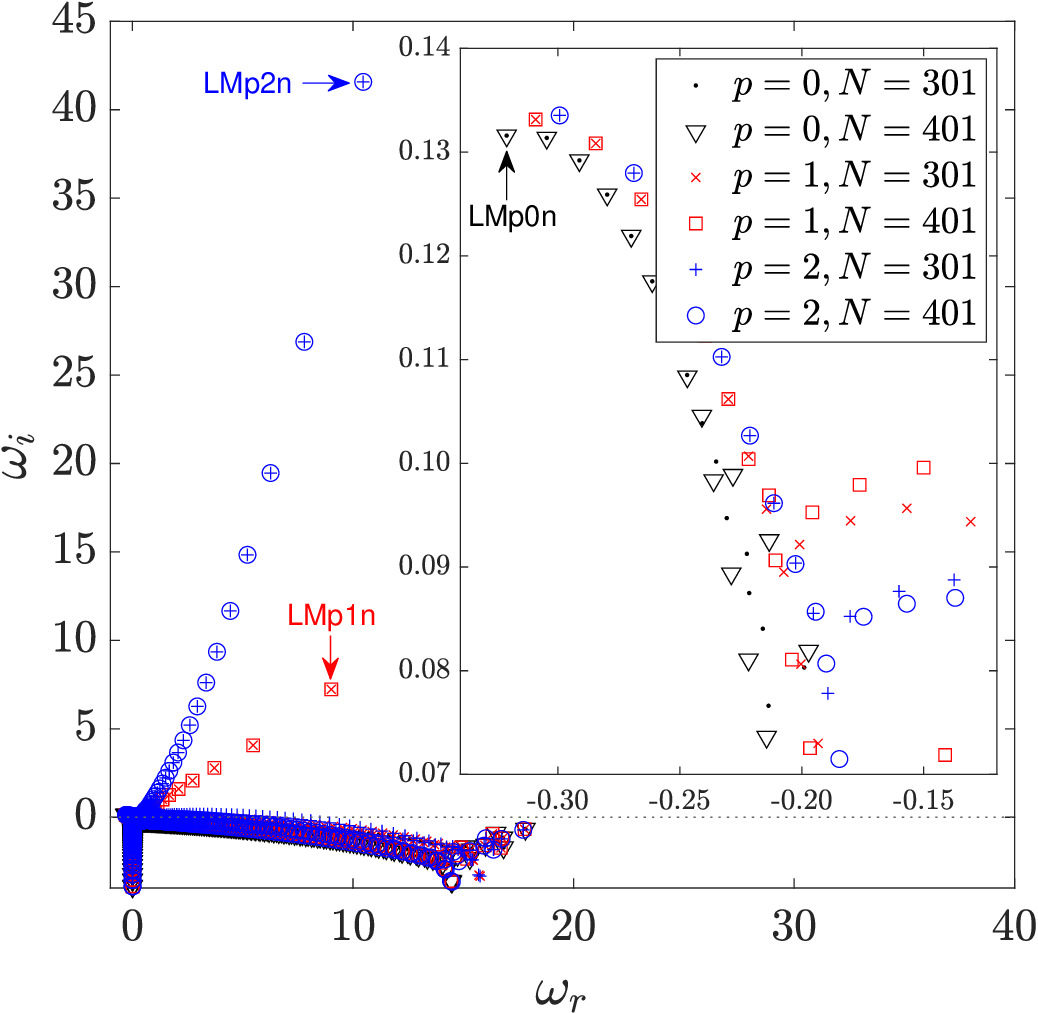}
	\put(-180,160){$(b)$}	
	\caption{Eigenspectra for different gravitational acceleration profiles ($g(r)=r_o^p/r^p$ with $p=0,1,2$, respectively) for the quasi-Keplerian flow at $(\eta,Pr,Ri,Ta)=(0.05,10^{-3},10^{-3},10^{16})$. Panel $(a)$ is for the case at $(k,m)=(75.3178,0)$; panel $(b)$ is for the case at $(k,m)=(80.4170,1)$; these two panels correspond to the two maxima respectively in figure \ref{Fig:eta005_Ta1e16_m01}$(a)$. Labels point to the leading modes for each $p$.}
	\label{Fig:eigenspectra_gr}
\end{figure}

\begin{figure}
	\centering
	\includegraphics[width=0.99\textwidth,trim= 30 0 30 0,clip]{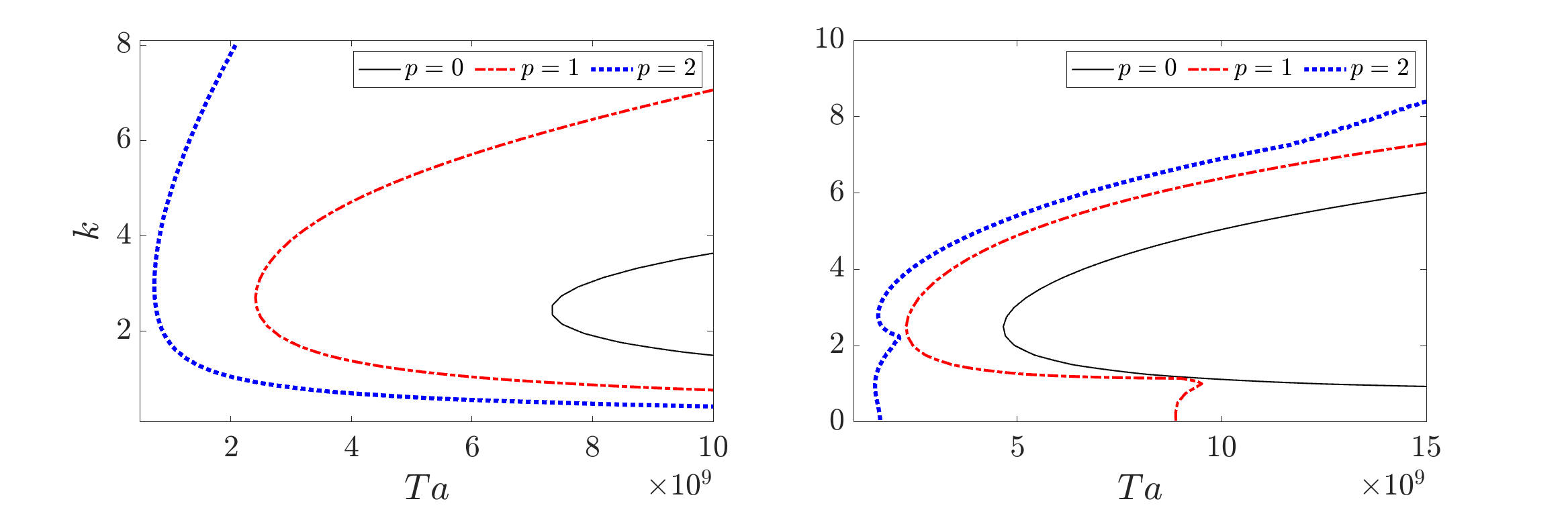}
	\put(-374,117){$(a)$}
	\put(-194,117){$(b)$}	
	\caption{Neutral curves for different gravitational acceleration profiles ($g(r)=r_o^p/r^p$ with $p=0,1,2$, respectively) for the quasi-Keplerian flow at $(\eta,Pr,Ri)=(0.05,10^{-3},10^{-3})$ and: $(a)$ $m=0$; $(b)$ $m=1$. The neutral curves for $p=1$ are exactly those in figure \ref{Fig:NC_eta005_m0123}$(a,b)$.}
	\label{Fig:NC_gr}
\end{figure}

Our calculations indicate that, with all the other parameters fixed, compared to the case with $p=1$, a stronger gravity field ($p=2$) is destabilising, while a weaker gravity field ($p=0$) is stabilising. This behaviour is illustrated by the eigenspectra shown in figure \ref{Fig:eigenspectra_gr}. Panel \ref{Fig:eigenspectra_gr}$(a)$ displays the axisymmetric linear instability for the flow at $(\eta,Pr,Ri,Ta)=(0.05,10^{-3},10^{-3},10^{16})$. The leading unstable modes for $p=0,1,2$ are labelled as ``LMp0a'', ``LMp1a'' and ``LMp2a'', respectively. Compared to mode ``LMp1a'', mode ``LMp2a'' exhibits a higher linear growth rate $\omega_i$, while mode ``LMp0a'' shows a lower growth rate. A similar trend is observed for non-axisymmetric modes in panel \ref{Fig:eigenspectra_gr}$(b)$.
Neutral curves provide another perspective on the destabilising or stabilising effects. As shown in figure \ref{Fig:NC_gr}$(a,b)$, the neutral curve for $p=2$ shifts to the left, indicating a smaller critical Taylor number and thus a destabilising effect due to the stronger gravity field. Conversely, for the constant profile ($p=0$), the neutral curve shifts to the right and results in a larger critical Taylor number, demonstrating the stabilising effect of a weaker gravity field.
It is worth noting that the linear critical Taylor numbers $Ta_c$ for the three gravitational acceleration profiles are of the same order of magnitude. This suggests that the constant gravitational acceleration profile with $p=0$, which is the only realisable profile in experiments to date, could be used to predict approximately the instability onset of the flow with $p=1$ in future experiments.

In the end, we mention in passing that the influence of the gravitational acceleration profile on the linear instability can be understood in terms of buoyancy. Increased gravity enhances the buoyancy force, which promotes flow instability, as evidenced by the gravitational buoyancy term in the governing equation \ref{eq2.1b}.


\bibliographystyle{jfm}
\bibliography{jfm}

\begin{thebibliography}{56}
\expandafter\ifx\csname natexlab\endcsname\relax\def\natexlab#1{#1}\fi
\def\au#1{#1} \def\ed#1{#1} \def\yr#1{#1}\def\at#1{#1}\def\jt#1{\textit{#1}}
  \def\bt#1{#1}\def\bvol#1{\textbf{#1}} \def\vol#1{#1} \def\pg#1{#1}
  \def\publ#1{#1}\def\arxiv#1{#1}\def\org#1{#1}\def\st#1{\textit{#1}}

\bibitem[Abramowicz \& Straub(2014)]{Abramowicz2014Accretion}
{\sc \au{Abramowicz, M.A.} \& \au{Straub, O.}} \yr{2014}  \at{{A}ccretion
  discs}.  \jt{Scholarpedia}  \bvol{9}~(8),  \pg{2408}, revision \#145813.

\bibitem[Ali \& Weidman(1990)]{Ali1990Stability}
{\sc \au{Ali, M.} \& \au{Weidman, P.D.}} \yr{1990}  \at{{On the stability of
  circular Couette flow with radial heating}}.  \jt{J. Fluid Mech.}
  \bvol{220},  \pg{53--84}.

\bibitem[Antoine {\em et~al.\/}(2023)Antoine, Martin, Vasyl \&
  Christoph]{Antoine2023Thermo}
{\sc \au{Antoine, M.}, \au{Martin, M.}, \au{Vasyl, M.} \& \au{Christoph, E.}}
  \yr{2023}  \at{Thermo-electric convection in a cylindrical annulus during a
  sounding rocket flight}.  \jt{J. Fluid Mech.}  \pg{p. 972}.

\bibitem[Armitage(2011)]{Armitage2011Dynamics}
{\sc \au{Armitage, P.J.}} \yr{2011}  \at{{Dynamics of protoplanetary disks}}.
  \jt{Annu. Rev. Astron. Astrophys.}  \bvol{49},  \pg{195--236}.

\bibitem[Avila(2012)]{Avila2012Stability}
{\sc \au{Avila, M.}} \yr{2012}  \at{{Stability and angular-momentum transport
  of fluid flows between corotating cylinders}}.  \jt{Phys. Rev. Lett.}
  \bvol{108}~(12),  \pg{124501}.

\bibitem[Balbus \& Hawley(1991)]{Balbus1991Powerful}
{\sc \au{Balbus, S.A.} \& \au{Hawley, J.F.}} \yr{1991}  \at{{A powerful local
  shear instability in weakly magnetized disks. I-Linear analysis. II-Nonlinear
  evolution}}.  \jt{Astrophys. J.}  \bvol{376},  \pg{214--233}.

\bibitem[Becker \& Kaye(1962)]{Becker1962Influence}
{\sc \au{Becker, K.M.} \& \au{Kaye, J.}} \yr{1962}  \at{{The influence of a
  radial temperature gradient on the instability of fluid flow in an annulus
  with an inner rotating cylinder}}.  \jt{J. Heat Transfer.}  \bvol{84}~(2),
  \pg{106--110}.

\bibitem[Bodenschatz {\em et~al.\/}(2000)Bodenschatz, Pesch \&
  Ahlers]{Bodenschatz2000Recent}
{\sc \au{Bodenschatz, E.}, \au{Pesch, W.} \& \au{Ahlers, G.}} \yr{2000}
  \at{{Recent developments in Rayleigh-B{\'e}nard convection}}.  \jt{Annu. Rev.
  Fluid Mech.}  \bvol{32}~(1),  \pg{709--778}.

\bibitem[Chiang \& Goldreich(1997)]{Chiang1997Spectral}
{\sc \au{Chiang, E.I.} \& \au{Goldreich, P.}} \yr{1997}  \at{{Spectral energy
  distributions of T Tauri stars with passive circumstellar disks}}.
  \jt{Astrophys. J.}  \bvol{490}~(1),  \pg{368}.

\bibitem[Deguchi(2017)]{Deguchi2017Linear}
{\sc \au{Deguchi, K.}} \yr{2017}  \at{{Linear instability in Rayleigh-stable
  Taylor-Couette flow}}.  \jt{Phys. Rev. E}  \bvol{95}~(2),  \pg{021102}.

\bibitem[Dubrulle {\em et~al.\/}(2005)Dubrulle, Dauchot, Daviaud, Longaretti,
  Richard \& Zahn]{Dubrulle2005Stability}
{\sc \au{Dubrulle, B.}, \au{Dauchot, O.}, \au{Daviaud, F.}, \au{Longaretti,
  P.-Y.}, \au{Richard, D.} \& \au{Zahn, J.-P.}} \yr{2005}  \at{{Stability and
  turbulent transport in Taylor--Couette flow from analysis of experimental
  data}}.  \jt{Phys. Fluids}  \bvol{17}~(9).

\bibitem[Ecke \& Shishkina(2023)]{Ecke2023Turbulent}
{\sc \au{Ecke, R.E.} \& \au{Shishkina, O.}} \yr{2023}  \at{{Turbulent rotating
  Rayleigh--b{\'e}nard convection}}.  \jt{Annu. Rev. Fluid Mech.}  \bvol{55},
  \pg{603--638}.

\bibitem[Eckhardt {\em et~al.\/}(2007)Eckhardt, Schneider, Hof \&
  Westerweel]{Eckhardt2007Turbulence}
{\sc \au{Eckhardt, B.}, \au{Schneider, T.M.}, \au{Hof, B.} \& \au{Westerweel,
  J.}} \yr{2007}  \at{{Turbulence transition in pipe flow}}.  \jt{Annu. Rev.
  Fluid Mech.}  \bvol{39},  \pg{447--468}.

\bibitem[Edlund \& Ji(2014)]{Edlund2014Nonlinear}
{\sc \au{Edlund, E.M.} \& \au{Ji, H.}} \yr{2014}  \at{{Nonlinear stability of
  laboratory quasi-Keplerian flows}}.  \jt{Phys. Rev. E}  \bvol{89}~(2),
  \pg{021004}.

\bibitem[Fleming \& Stone(2003)]{Fleming2003Local}
{\sc \au{Fleming, T.} \& \au{Stone, J.M.}} \yr{2003}  \at{{Local
  magnetohydrodynamic models of layered accretion disks}}.  \jt{Astrophys. J.}
  \bvol{585}~(2),  \pg{908}.

\bibitem[Fromang \& Lesur(2019)]{Fromang2019Angular}
{\sc \au{Fromang, S.} \& \au{Lesur, G.}} \yr{2019}  \at{{Angular momentum
  transport in accretion disks: a hydrodynamical perspective}}.  \jt{EAS
  Publications Series}  \bvol{82},  \pg{391--413}.

\bibitem[Gammie(1996)]{Gammie1996Layered}
{\sc \au{Gammie, C.F.}} \yr{1996}  \at{{Layered accretion in T Tauri disks}}.
  \jt{Astrophys. J.}  \bvol{457},  \pg{355}.

\bibitem[Garg {\em et~al.\/}(2018)Garg, Chaudhary, Khalid, Shankar \&
  Subramanian]{Garg2018Viscoelastic}
{\sc \au{Garg, P.}, \au{Chaudhary, I.}, \au{Khalid, M.}, \au{Shankar, V.} \&
  \au{Subramanian, G.}} \yr{2018}  \at{{Viscoelastic pipe flow is linearly
  unstable}}.  \jt{Phys. Rev. Lett.}  \bvol{121}~(2),  \pg{024502}.

\bibitem[Grossmann(2000)]{Grossmann2000Onset}
{\sc \au{Grossmann, S.}} \yr{2000}  \at{{The onset of shear flow turbulence}}.
  \jt{Rev. Mod. Phys.}  \bvol{72}~(2),  \pg{603}.

\bibitem[Grossmann {\em et~al.\/}(2016)Grossmann, Lohse \&
  Sun]{Grossmann2016High}
{\sc \au{Grossmann, S.}, \au{Lohse, D.} \& \au{Sun, C.}} \yr{2016}
  \at{{High--Reynolds number Taylor-Couette turbulence}}.  \jt{Annu. Rev. Fluid
  Mech.}  \bvol{48},  \pg{53--80}.

\bibitem[Held \& Latter(2018)]{Held2018Hydrodynamic}
{\sc \au{Held, L.E.} \& \au{Latter, H.N.}} \yr{2018}  \at{{Hydrodynamic
  convection in accretion discs}}.  \jt{Mon. Not. R. Astron. Soc.}
  \bvol{480}~(4),  \pg{4797--4816}.

\bibitem[Hollerbach \& R{\"u}diger(2005)]{Hollerbach2005New}
{\sc \au{Hollerbach, R.} \& \au{R{\"u}diger, G.}} \yr{2005}  \at{{New type of
  magnetorotational instability in cylindrical Taylor-Couette flow}}.
  \jt{Phys. Rev. Lett.}  \bvol{95}~(12),  \pg{124501}.

\bibitem[Ikoma \& Hori(2012)]{Ikoma2012Situ}
{\sc \au{Ikoma, M.} \& \au{Hori, Y.}} \yr{2012}  \at{{In situ accretion of
  hydrogen-rich atmospheres on short-period super-Earths: implications for the
  Kepler-11 planets}}.  \jt{Astrophys. J.}  \bvol{753}~(1),  \pg{66}.

\bibitem[Jenny \& Nsom(2007)]{Jenny2007Primary}
{\sc \au{Jenny, M.} \& \au{Nsom, B.}} \yr{2007}  \at{{Primary instability of a
  Taylor-Couette flow with a radial stratification and radial buoyancy}}.
  \jt{Phys. Fluids}  \bvol{19}~(10).

\bibitem[Ji \& Balbus(2013)]{Ji2013Angular}
{\sc \au{Ji, H.} \& \au{Balbus, S.}} \yr{2013}  \at{{Angular momentum transport
  in astrophysics and in the lab}}.  \jt{Phys. Today}  \bvol{66}~(8),
  \pg{27--33}.

\bibitem[Ji {\em et~al.\/}(2006)Ji, Burin, Schartman \&
  Goodman]{Ji2006Hydrodynamic}
{\sc \au{Ji, H.}, \au{Burin, M.}, \au{Schartman, E.} \& \au{Goodman, J.}}
  \yr{2006}  \at{{Hydrodynamic turbulence cannot transport angular momentum
  effectively in astrophysical disks}}.  \jt{Nature}  \bvol{444}~(7117),
  \pg{343--346}.

\bibitem[Jiang {\em et~al.\/}(2022)Jiang, Wang, Liu \&
  Sun]{Jiang2022Experimental}
{\sc \au{Jiang, H.}, \au{Wang, D.}, \au{Liu, S.} \& \au{Sun, C.}} \yr{2022}
  \at{Experimental evidence for the existence of the ultimate regime in rapidly
  rotating turbulent thermal convection}.  \jt{Phys. Rev. Lett.}
  \bvol{129}~(20),  \pg{204502}.

\bibitem[Jiang {\em et~al.\/}(2020)Jiang, Zhu, Wang, Huisman \&
  Sun]{Jiang2020Supergravitational}
{\sc \au{Jiang, H.}, \au{Zhu, X.}, \au{Wang, D.}, \au{Huisman, S.~G.} \&
  \au{Sun, C.}} \yr{2020}  \at{Supergravitational turbulent thermal
  convection}.  \jt{Sci. Adv.}  \bvol{6}~(40),  \pg{eabb8676}.

\bibitem[Kang {\em et~al.\/}(2017)Kang, Meyer, Mutabazi \&
  Yoshikawa]{Kang2017Radial}
{\sc \au{Kang, C.}, \au{Meyer, A.}, \au{Mutabazi, I.} \& \au{Yoshikawa, H.N.}}
  \yr{2017}  \at{{Radial buoyancy effects on momentum and heat transfer in a
  circular Couette flow}}.  \jt{Phys. Rev. Fluids}  \bvol{2}~(5),  \pg{053901}.

\bibitem[Klahr \& Hubbard(2014)]{Klahr2014Convective}
{\sc \au{Klahr, H.} \& \au{Hubbard, A.}} \yr{2014}  \at{{Convective
  overstability in radially stratified accretion disks under thermal
  relaxation}}.  \jt{Astrophys. J.}  \bvol{788}~(1),  \pg{21}.

\bibitem[Latter(2016)]{Latter2016Convective}
{\sc \au{Latter, H.N.}} \yr{2016}  \at{{On the convective overstability in
  protoplanetary discs}}.  \jt{Mon. Not. R. Astron. Soc.}  \bvol{455}~(3),
  \pg{2608--2618}.

\bibitem[Leclercq {\em et~al.\/}(2016)Leclercq, Partridge, Augier, Caulfield,
  B. \& Linden]{Leclercq2016Nonlinear}
{\sc \au{Leclercq, C.}, \au{Partridge, J.~L.}, \au{Augier, P.}, \au{Caulfield,
  C. C.~P.}, \au{B., Dalziel~S.} \& \au{Linden, P.~F.}} \yr{2016}
  \at{{Nonlinear waves in stratified Taylor--Couette flow. Part 1. Layer
  formation}}.  \jt{arXiv e-prints} .

\bibitem[Lesur {\em et~al.\/}(2022)Lesur, Ercolano, Flock, Lin, Yang, Barranco,
  Benitez-Llambay, Goodman, Johansen, Klahr, Laibe, Lyra, Marcus, Nelson,
  Squire, Simon, Turner, Umurhan \& Youdin]{lesur2022}
{\sc \au{Lesur, G.}, \au{Ercolano, B.}, \au{Flock, M.}, \au{Lin, M.~K.},
  \au{Yang, C.~C.}, \au{Barranco, J.~A.}, \au{Benitez-Llambay, P.},
  \au{Goodman, J.}, \au{Johansen, A.}, \au{Klahr, H.}, \au{Laibe, G.},
  \au{Lyra, W.}, \au{Marcus, P.}, \au{Nelson, R.~P.}, \au{Squire, J.},
  \au{Simon, J.~B.}, \au{Turner, N.}, \au{Umurhan, O.~M.} \& \au{Youdin,
  A.~N.}} \yr{2022}  \at{Hydro-, magnetohydro-, and dust-gas dynamics of
  protoplanetary disks}.  \jt{arXiv 2203.09821} .

\bibitem[Lesur \& Ogilvie(2010)]{Lesur2010Angular}
{\sc \au{Lesur, G.} \& \au{Ogilvie, G.I.}} \yr{2010}  \at{On the angular
  momentum transport due to vertical convection in accretion discs}.  \jt{Mon.
  Not. R. Astron. Soc. Lett.}  \bvol{404}~(1),  \pg{L64--L68}.

\bibitem[Lesur \& Papaloizou(2010)]{Lesur2010Subcritical}
{\sc \au{Lesur, G.} \& \au{Papaloizou, J.C.B.}} \yr{2010}  \at{{The subcritical
  baroclinic instability in local accretion disc models}}.  \jt{Astron.
  Astrophys.}  \bvol{513},  \pg{A60}.

\bibitem[Lyra(2014)]{Lyra2014Convective}
{\sc \au{Lyra, W.}} \yr{2014}  \at{{Convective overstability in accretion
  disks: Three-dimensional linear analysis and nonlinear saturation}}.
  \jt{Astrophys. J.}  \bvol{789}~(1),  \pg{77}.

\bibitem[Maslowe(1986)]{Maslowe1986}
{\sc \au{Maslowe, S~A}} \yr{1986}  \at{Critical layers in shear flows}.
  \jt{Annu. Rev. Fluid Mech.}  \bvol{18},  \pg{405--432}.

\bibitem[McKeon(2017)]{McKeon2017}
{\sc \au{McKeon, B.~J.}} \yr{2017}  \at{The engine behind (wall) turbulence:
  perspectives on scale interactions}.  \jt{Journal of Fluid Mechanics}
  \bvol{817},  \pg{P1}.

\bibitem[Meyer {\em et~al.\/}(2021)Meyer, Mutabazi \&
  Yoshikawa]{Meyer2021Stability}
{\sc \au{Meyer, A.}, \au{Mutabazi, I.} \& \au{Yoshikawa, H.N.}} \yr{2021}
  \at{{Stability of Rayleigh-stable Couette flow between two differentially
  heated cylinders}}.  \jt{Phys. Rev. Fluids}  \bvol{6}~(3),  \pg{033905}.

\bibitem[Meyer {\em et~al.\/}(2015)Meyer, Yoshikawa \&
  Mutabazi]{Meyer2015Effect}
{\sc \au{Meyer, A.}, \au{Yoshikawa, H.N.} \& \au{Mutabazi, I.}} \yr{2015}
  \at{{Effect of the radial buoyancy on a circular Couette flow}}.  \jt{Phys.
  Fluids}  \bvol{27}~(11).

\bibitem[Mineshige(1993)]{Mineshige1993Accretion}
{\sc \au{Mineshige, S.}} \yr{1993}  \at{{Accretion disk instabilities}}.
  \jt{IAU Colloquia.}  \bvol{134},  \pg{83--103}.

\bibitem[Mishra {\em et~al.\/}(2024)Mishra, Mamatsashvili, Seilmayer \&
  Stefani]{Mishra2024One}
{\sc \au{Mishra, Ashish}, \au{Mamatsashvili, George}, \au{Seilmayer, Martin} \&
  \au{Stefani, Frank}} \yr{2024}  \at{One-winged butterflies: mode selection
  for azimuthal magnetorotational instability by thermal convection}.
  \jt{arXiv preprint arXiv:2403.09764} .

\bibitem[Nelson {\em et~al.\/}(2013)Nelson, Gressel \&
  Umurhan]{Nelson2013Linear}
{\sc \au{Nelson, R.P.}, \au{Gressel, O.} \& \au{Umurhan, O.M.}} \yr{2013}
  \at{{Linear and non-linear evolution of the vertical shear instability in
  accretion discs}}.  \jt{Mon. Not. R. Astron. Soc}  \bvol{435}~(3),
  \pg{2610--2632}.

\bibitem[Ostilla-M{\'o}nico {\em et~al.\/}(2014)Ostilla-M{\'o}nico, Verzicco,
  Grossmann \& Lohse]{Ostilla2014Turbulence}
{\sc \au{Ostilla-M{\'o}nico, R.}, \au{Verzicco, R.}, \au{Grossmann, S.} \&
  \au{Lohse, D.}} \yr{2014}  \at{{Turbulence decay towards the linearly stable
  regime of Taylor--Couette flow}}.  \jt{J. Fluid Mech.}  \bvol{748},  \pg{R3}.

\bibitem[Pringle(1981)]{Pringle1981Accretion}
{\sc \au{Pringle, J.E.}} \yr{1981}  \at{{Accretion discs in astrophysics}}.
  \jt{Annu. Rev. Astron. Astrophys.}  \bvol{19}~(1),  \pg{137--160}.

\bibitem[Rayleigh(1917)]{Rayleigh1917Dynamics}
{\sc \au{Rayleigh, L.}} \yr{1917}  \at{{On the dynamics of revolving fluids}}.
  \jt{Proceedings of the Royal Society of London. Series A, Containing Papers
  of a Mathematical and Physical Character}  \bvol{93}~(648),  \pg{148--154}.

\bibitem[R{\"u}diger {\em et~al.\/}(2003)R{\"u}diger, Schultz \&
  Shalybkov]{Rudiger2003Linear}
{\sc \au{R{\"u}diger, G.}, \au{Schultz, M.} \& \au{Shalybkov, D.}} \yr{2003}
  \at{{Linear magnetohydrodynamic Taylor-Couette instability for liquid
  sodium}}.  \jt{Phys. Rev. E}  \bvol{67}~(4),  \pg{046312}.

\bibitem[Seilmayer {\em et~al.\/}(2020)Seilmayer, Ogbonna \&
  Stefani]{Seilmayer2020Convection}
{\sc \au{Seilmayer, M.}, \au{Ogbonna, J.} \& \au{Stefani, F.}} \yr{2020}
  \at{{Convection-caused symmetry breaking of azimuthal magnetorotational
  instability in a liquid metal Taylor-Couette Flow}}.
  \jt{Magnetohydrodynamics}  \bvol{56}~(2/3),  \pg{225 -- 235}.

\bibitem[Shakura \& Sunyaev(1973)]{Shakura1973Black}
{\sc \au{Shakura, N.I.} \& \au{Sunyaev, R.A.}} \yr{1973}  \at{{Black holes in
  binary systems. Observational appearance.}}  \jt{Astron. Astrophys.}
  \bvol{24},  \pg{337--355}.

\bibitem[Shi {\em et~al.\/}(2017)Shi, Hof, Rampp \& Avila]{Shi2017Hydrodynamic}
{\sc \au{Shi, L.}, \au{Hof, B.}, \au{Rampp, M.} \& \au{Avila, M.}} \yr{2017}
  \at{{Hydrodynamic turbulence in quasi-Keplerian rotating flows}}.  \jt{Phys.
  Fluids}  \bvol{29}~(4).

\bibitem[Trefethen(2000)]{Trefethen2000Spectral}
{\sc \au{Trefethen, L.N.}} \yr{2000} {\em {Spectral methods in MATLAB}\/}.
  \publ{SIAM}.

\bibitem[Velikhov(1959)]{Velikhov1959Stability}
{\sc \au{Velikhov, E.}} \yr{1959}  \at{Stability of an ideally conducting
  liquid flowing between cylinders rotating in a magnetic field}.  \jt{J. Exp.
  Theor. Phys.}  \bvol{9},  \pg{995--998}.

\bibitem[Wang {\em et~al.\/}(2022{\natexlab{{\em a\/}}})Wang, Jiang, Liu, Zhu
  \& Sun]{Wang2022Effects}
{\sc \au{Wang, D.}, \au{Jiang, H.}, \au{Liu, S.}, \au{Zhu, X.} \& \au{Sun, C.}}
  \yr{2022{\natexlab{{\em a\/}}}}  \at{{Effects of radius ratio on annular
  centrifugal Rayleigh--B{\'e}nard convection}}.  \jt{J. Fluid Mech.}
  \bvol{930},  \pg{A19}.

\bibitem[Wang {\em et~al.\/}(2022{\natexlab{{\em b\/}}})Wang, Gilson, Ebrahimi,
  Goodman, Caspary, Winarto \& Ji]{Wang2022Identification}
{\sc \au{Wang, Y.}, \au{Gilson, E.P.}, \au{Ebrahimi, F.}, \au{Goodman, J.},
  \au{Caspary, K.J.}, \au{Winarto, H.W.} \& \au{Ji, H.}}
  \yr{2022{\natexlab{{\em b\/}}}}  \at{{Identification of a non-axisymmetric
  mode in laboratory experiments searching for standard magnetorotational
  instability}}.  \jt{Nature Comm.}  \bvol{13}~(1),  \pg{4679}.

\bibitem[Wang {\em et~al.\/}(2022{\natexlab{{\em c\/}}})Wang, Gilson, Ebrahimi,
  Goodman \& Ji]{Wang2022Observation}
{\sc \au{Wang, Y.}, \au{Gilson, E.P.}, \au{Ebrahimi, F.}, \au{Goodman, J.} \&
  \au{Ji, H.}} \yr{2022{\natexlab{{\em c\/}}}}  \at{{Observation of
  axisymmetric standard magnetorotational instability in the laboratory}}.
  \jt{Phys. Rev. Lett.}  \bvol{129}~(11),  \pg{115001}.

\bibitem[Yavneh {\em et~al.\/}(2001)Yavneh, Mcwilliams \&
  Molemaker]{Yavneh2001Non}
{\sc \au{Yavneh, I.}, \au{Mcwilliams, J.~C.} \& \au{Molemaker, M.~J.}}
  \yr{2001}  \at{{Non-axisymmetric instability of centrifugally stable
  stratified Taylor–Couette flow}}.  \jt{J. Fluid Mech.}  \bvol{448}.

\end{thebibliography}



\end{document}